\documentclass[aps,twocolumn,prd,superscriptaddress,preprintnumbers]{revtex4-2}

\usepackage[utf8]{inputenc}

\usepackage{mathtools}
\usepackage{amsfonts}
\usepackage{mathrsfs}
\usepackage{bbm}
\usepackage{slashed}
\usepackage{tensor}

\usepackage{graphicx}
\usepackage{array}

\usepackage{placeins}
\usepackage{makecell}
\usepackage[caption=false]{subfig}
\usepackage{float}

\usepackage{xspace}
\usepackage{xfrac}
\usepackage{hyperref}
\usepackage[nameinlink]{cleveref}
\usepackage{appendix}
\usepackage{units}

\usepackage{xifthen}
\usepackage{booktabs}
\usepackage{multirow}
\usepackage{courier}
\usepackage[dvipsnames]{xcolor}
\hypersetup{
	colorlinks,
	linkcolor={blue!75!black},
	citecolor={blue!75!black},
	urlcolor={blue!75!black}
}

\newcommand*{\eg}{e.g.\@\xspace}
\newcommand*{\ie}{i.e.\@\xspace}

\graphicspath{{./figures/}}

\newcommand{\gettitle}{Yang-Mills glueball masses from spectral reconstruction}

\newcommand{\getHeidelbergAffiliation}{\affiliation{Institut f\"ur Theoretische Physik,	Universit\"at Heidelberg, Philosophenweg 16, D-69120 Heidelberg, Germany}}
\newcommand{\getDarmstadtAffiliation}{\affiliation{Institut f\"ur Kernphysik, Technische Universit\"at Darmstadt, D-64289 Darmstadt, Germany}}
\newcommand{\getEMMIAffiliation}{\affiliation{ExtreMe Matter Institute EMMI, GSI, Planckstr. 1, 64291 Darmstadt, Germany}}
\newcommand{\getMITAffiliation}{\affiliation{Center for Theoretical Physics, Massachusetts Institute of Technology, Cambridge, MA 02139, USA}}
\newcommand{\getIAIFIAffiliation}{\affiliation{The NSF AI Institute for Artificial Intelligence and Fundamental Interactions}}

\hypersetup{
	pdftitle={\gettitle},
	pdfauthor={Pawlowski, Schneider, Turnwald, Urban, Wink},
	pdfkeywords={Yang-Mills theory}
		{correlation functions}
		{functional renormalisation group}
		{glueballs}
		{spectral reconstruction}
		{gaussian process regression},
	bookmarksopen=true,
	bookmarksopenlevel=2,
	bookmarksnumbered=true
}

\begin{document}

\title{\gettitle}

\author{Jan M. Pawlowski}
\getHeidelbergAffiliation
\getEMMIAffiliation

\author{Coralie S. Schneider}
\getHeidelbergAffiliation

\author{Jonas Turnwald}
\email{j.turnwald@theorie.ikp.physik.tu-darmstadt.de}
\getHeidelbergAffiliation
\getDarmstadtAffiliation

\author{Julian M. Urban}
\getMITAffiliation
\getIAIFIAffiliation

\author{Nicolas~Wink}
\getDarmstadtAffiliation

\preprint{MIT-CTP/5502}

\begin{abstract}
We compute masses of the two lightest glueballs from spectral reconstructions of timelike interaction channels of the four-gluon vertex in Landau gauge Yang-Mills theory. The Euclidean spacelike dressings of the vertex are calculated with the functional renormalisation group. For the spectral reconstruction of these Euclidean data, we employ Gaussian process regression. The glueball resonances can be identified straightforwardly and we obtain $m_{sc} = 1870(75)\,$MeV as well as $m_{ps} = 2700(120)\,$MeV, in accordance with functional bound state and lattice calculations.
\end{abstract}

\maketitle

\section{Introduction}\label{sec:Introduction}

The hadronic spectrum of Yang-Mills theory and QCD includes purely gluonic bound state contributions, the glueballs. The experimental verification of their existence is an important test of QCD; however, it is not yet conclusive~\cite{Klempt:2007cp, Crede:2008vw, Ochs:2013gi, Klempt:2022ipu} as these states are difficult to access due to their large overlap with other hadronic resonances. Different possible experimental candidates have been proposed, including the various $f_0$ states, some of which are expected to appear in decay channels of $J/\psi$~\cite{Sarantsev:2021ein, Klempt:2021wpg}. The overlap with other states also complicates their theoretical determination when considering QCD; for corresponding lattice calculations see~\cite{Chen:2021dvn, Gregory:2012hu, Brett:2019tzr}. In Yang-Mills theory, the situation is much simpler and the first few lightest states are well known; for lattice results see \eg~\cite{Morningstar:1999rf, Bali:1993fb, Chen:2005mg, Gregory:2012hu, Athenodorou:2020ani, Sakai:2022zdc}. For computations with functional approaches---in particular with a combination of Dyson-Schwinger equations (DSE) and Bethe-Salpeter equations (BSE)---see \eg~\cite{Meyers:2012ka, Sanchis-Alepuz:2015hma, Souza:2019ylx, Kaptari:2020qlt, Huber:2020ngt, Huber:2021yfy}.

In this work, we put forward a self-consistent functional ansatz for computing masses of bound states by exploiting their overlap with resonant interaction channels of gauge-fixed correlation functions. The approach is then used to determine the masses of the scalar (\mbox{$J^{PC} = 0^{++}$}) and pseudo-scalar (\mbox{$J^{PC} = 0^{-+}$}) Yang-Mills glueballs, utilising the fact that these states have overlap with channels of the four-gluon vertex that carry the respective symmetries, where they appear as peaks of the corresponding spectral functions. We use Gaussian process regression (GPR) to compute these spectral functions by reconstructing Euclidean correlators obtained within the functional renormalisation group (fRG) framework in~\cite{Pawlowski:2022oyq}. The inversion of the spectral representation is an ill-conditioned problem; see \eg~\cite{Cuniberti:2001hm, Burnier:2011jq, Shi:2022yqw}. The applicability of GPR to such linear inverse problems was discussed in \cite{valentine2020gaussian} and the approach has since been employed to compute ghost and gluon spectral functions from 2+1 flavour lattice QCD results~\cite{Horak:2021syv}.

This paper is organised as follows. In \Cref{sec:YM+SpecRep}, we introduce the spectral representation of Euclidean dressing functions and discuss the projections onto the four-point vertices in Yang-Mills theory. The reconstruction approach using GPR is described in \Cref{sec:GP:indirect}. In \Cref{sec:Results}, the resulting spectral functions are presented and we report the masses of the scalar and pseudo-scalar glueballs. We conclude in \Cref{sec:Conclusion}.

\section{Spectral representations of Yang-Mills correlation functions}\label{sec:YM+SpecRep}

Non-perturbative calculations of correlation functions in Yang-Mills theory are generally only possible in Euclidean space-time, either on the lattice or with functional approaches. While the latter framework in principle also allows direct access to real-time properties---albeit with a qualitatively increased  effort---real-time lattice calculations are often faced with intractable signal-to-noise problems. Accordingly, computing timelike observables such as transport coefficients, pole masses, and decay rates typically requires the reconstruction of timelike correlation functions from their spacelike Euclidean counterparts via the associated spectral representations.

\begin{figure*}
	\centering
	\includegraphics[width=.9\linewidth,trim=10 10 10 10,clip]{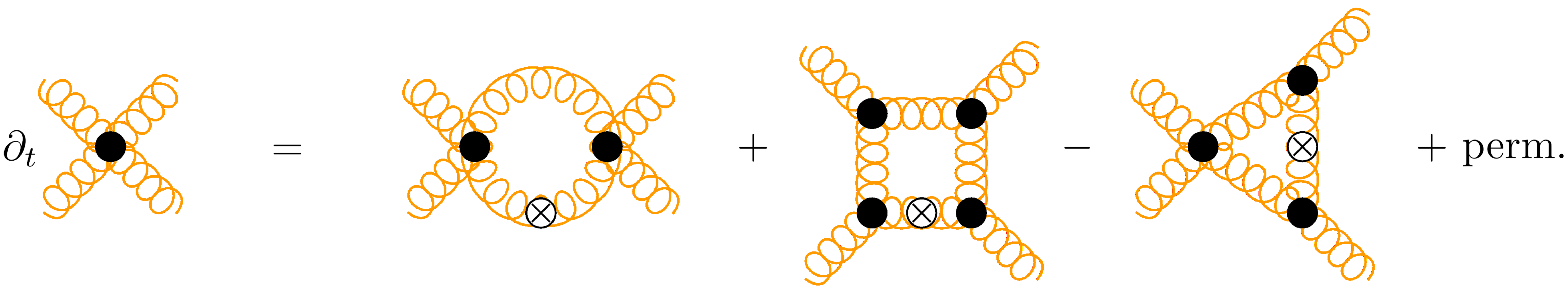}
	\caption[]{fRG equation for the s-channel four-gluon vertex dressing. Wiggly orange lines correspond to fully dressed gluon propagators; black dots indicate fully dressed vertices. Permutations include the various possible configurations of external legs as well as permutations of the regulator insertion (indicated by a crossed circle).}
	\label{fig:fRGGlueballs}
\end{figure*}

\subsection{Spectral representations}\label{sec:SpecRep}

Correlation functions of physical states, and in particular the two-point functions, admit a spectral representation. For the propagator---the inverse 1PI two-point function---this is the K{\"a}ll{\'e}n-Lehmann (KL) representation,
\begin{equation}\label{eq:KL}
	G(p) = \int_{0}^{\infty}\frac{d \omega}{2\pi}\frac{2 \omega\, \rho(\omega)}{p^2+\omega^2} \,.
\end{equation}
Here, $G(p)$ denotes the Euclidean propagator and $\rho(\omega)$ the spectral function, which is obtained by 
\begin{equation}\label{eq:anaCont}
	\rho(\omega) = 2 \lim\limits_{\epsilon\rightarrow 0^+}\text{Im}\;G(-i(\omega+i\epsilon)) \,.
\end{equation}

The spectral functions of asymptotic states are positive semi-definite and admit the interpretation of a probability density. In gauge theories, however, the situation becomes more complicated. To begin with, even the existence of a KL representation is not settled for ghost and gluon propagators, and \labelcref{eq:KL} may feature additional structures in the complex momentum plane; for a detailed discussion see~\cite{Pawlowski:2022oyq}. Moreover, in the Landau gauge, the gluon and ghost spectral functions exhibit negative infrared (IR) and ultraviolet (UV) tails; see \cite{Cyrol:2018xeq}. These properties can be inferred from the respective IR and UV asymptotic behaviour of the Euclidean correlation functions~\cite{Cyrol:2018xeq, Bonanno:2021squ, Horak:2021pfr, Pawlowski:2022oyq}, and these relations also hold true for the present analysis involving four-gluon vertices. Note also that while gauge-fixed correlation functions may not permit a KL representation, the scattering matrix elements are directly constructed in terms of these correlators and obey \labelcref{eq:KL}. Hence, features which are in direct correspondence to observables---such as bound states---can still be extracted from such gauge-fixed correlation functions.

\subsection{Four-gluon correlation function}\label{sec:4gluon}

\begin{figure*}[t]
	\centering
	\subfloat[]{%
		\includegraphics[width=0.5\linewidth]{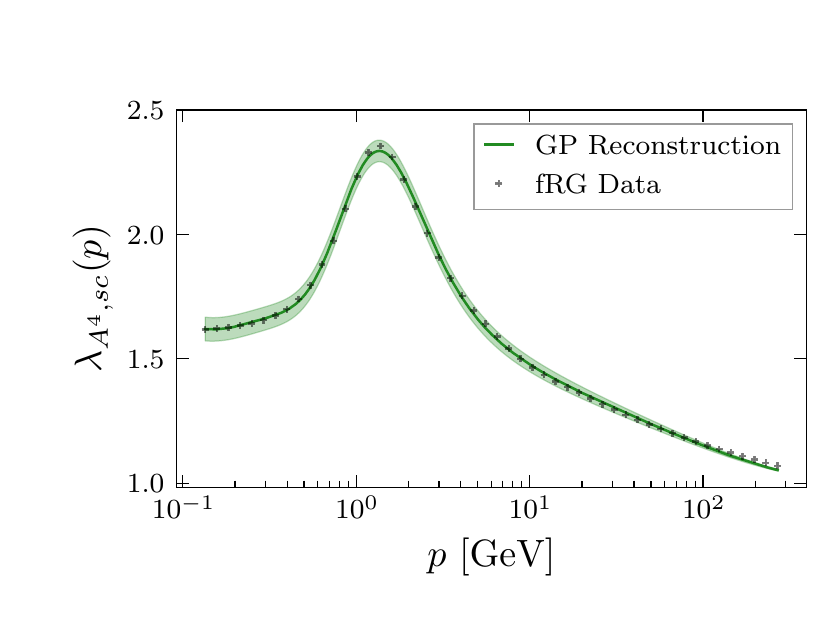}\label{fig:scalardressing}}%
	\subfloat[]{%
		\includegraphics[width=0.5\linewidth]{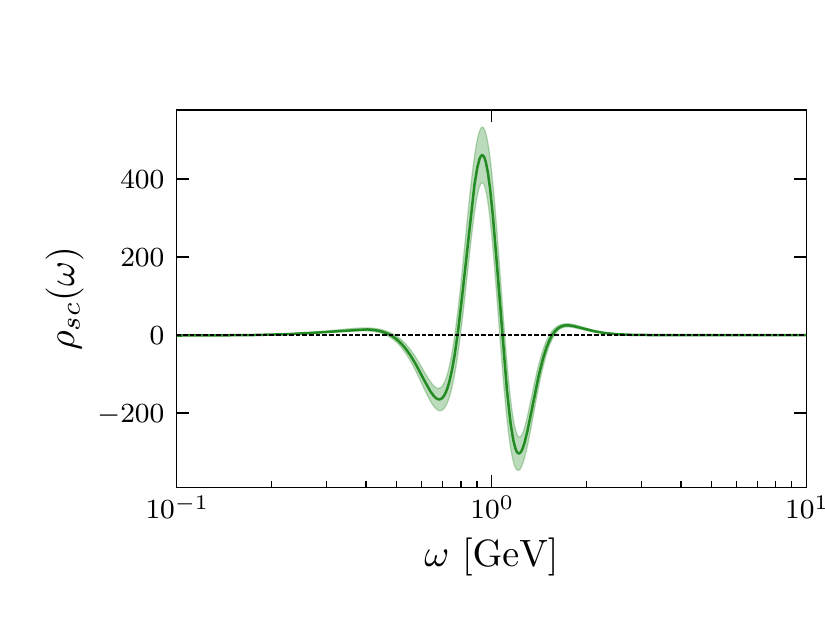}\label{fig:scalarsf}}
	\caption{(a) Euclidean dressing of the four-gluon vertex $\lambda_{A^4,sc}$ with the projection to obtain the scalar glueball mass, see \Cref{sec:4gluon}, from the fRG (black crosses). This is compared to the reconstruction from the GP (green line). The corresponding spectral function $\rho_{A^4,sc}$ over frequency $\omega$ obtained with GPR is shown in (b). The light green band represents the $1\sigma$ region.}
	\label{fig:scalarplot}
\end{figure*}

In the present work, we consider single interaction channels that have overlap with the bound states of interest. The spectral representations of these channels follow directly from the structure of the full, analytically continued correlation functions; see \eg~\cite{Evans:1991ky}. Up to minor modifications, they are given by~\labelcref{eq:KL}: for the relevant scalar dressings of the four-gluon vertex, we use~\cite{Horak:2020eng}
\begin{equation}\label{eq:KL4}
	\lambda_{A^4}(p^2) = \lambda_{A^4,0} +\int_0^{\infty} \frac{\text{d} \omega}{\pi}\frac{\omega \rho_{A^4}(\omega)}{p^2+\omega^2} \,.
\end{equation}
The constant part $\lambda_{A^4,0}$ accounts for the classical contribution. The Euclidean dressings of the interaction channels are computed with the fRG; for a recent review see~\cite{Dupuis:2020fhh}. The diagrammatic representation of the associated equation is shown in \Cref{fig:fRGGlueballs}; more details on the fRG approach and the specific computation for the vertex are deferred to \Cref{app:NumDetails}.

We remark in this context that correlation functions in Landau gauge Yang-Mills theory computed within sophisticated truncations to the fRG pass all available lattice benchmark tests; see \cite{Cyrol:2016tym, Pawlowski:2022oyq}. This concerns in particular the ghost and gluon propagators, whereas lattice results for vertices still exhibit large uncertainties. Nevertheless, since state-of-the-art functional results for correlation functions fully agree with lattice calculations within statistical errors, any reconstruction based on the former approaches is consistent with the latter. 

In order to access the masses of the scalar $J^{PC}= 0^{++}$ and pseudo-scalar $J^{PC}= 0^{-+}$ glueball, we have to determine tensor structures and momentum channels that overlap with these states. In general, it is desirable that the chosen channels have overlap only with the states of interest, as any reconstruction method faces increasing problems with multi-peak structures due to the exponential suppression of heavier states in the Euclidean data. Accordingly, their resolution requires an exponentially increasing accuracy, contributing to the ill-conditioned nature of the reconstruction problem.

For the scalar glueball, the above requirement is particularly simple to satisfy, since it is the lightest excitation and the classical tensor structure suffices, \ie
\begin{equation}\label{eq:classical_tensor}
\begin{aligned}
	\tau_{\text{\tiny cl}, \mu\nu\rho\sigma}^{abcd}
	&= f^{abe}f^{cde}(\delta_{\mu\rho} \delta_{\nu\sigma} - \delta_{\mu\sigma} \delta_{\nu\rho}) \\
	&+ f^{ace}f^{bde}(\delta_{\mu\nu} \delta_{\rho\sigma} - \delta_{\mu\sigma} \delta_{\nu\rho})  \\
	&+ f^{ade}f^{bce}(\delta_{\mu\nu} \delta_{\rho\sigma} - \delta_{\mu\rho} \delta_{\nu\sigma})  \,.
\end{aligned}
\end{equation}
Correspondingly, we use
\begin{equation}\label{eq:pseudotensor}
\begin{aligned}
    \tau_{ps,\mu\nu\rho\sigma}^{abcd}(p_1,p_2) = &\frac{\epsilon_{\mu\nu\alpha\beta}p_1^\alpha p_2^\beta}{\sqrt{p_1^2 p_2^2}} \frac{\epsilon_{\rho\sigma\gamma\delta}p_1^\gamma p_2^\delta}{\sqrt{p_1^2 p_2^2}} \\[1ex]
    &(\delta^{ab}\delta^{cd} + \delta^{ac}\delta^{bd} + \delta^{ad}\delta^{bc})
\end{aligned}
\end{equation}
for the pseudo-scalar glueball, which has the correct transformation properties (see \eg~\cite{Meyers:2012ka, Huber:2020ngt}), and does not overlap with the scalar glueball. In \labelcref{eq:pseudotensor}, $\varepsilon_{\mu\nu\rho\sigma}$ denotes the fully antisymmetric tensor and the momenta are chosen to be orthogonal, $p_1\cdot p_2=0$.

Finally, we have to specify the momentum channels: we restrict ourselves to a single exchange momentum and the external (incoming and outgoing) momenta are chosen to have the same magnitude, $p^2 \equiv p_1^2 = p_2^2$. This leaves us with two invariants: $p^2$ and $x=(p_1\cdot p_2)/p^2$. For the scalar glueball, the momenta are chosen to be parallel, \ie~$x=1$; for the pseudo-scalar one, they are chosen to be orthogonal, \ie~$x=0$. Further details on the projection operators are given in \Cref{app:Proj}.

\section{Gaussian process regression with indirect observations}\label{sec:GP:indirect}

GPR is widely employed as a non-parametric interpolation method for noisy observations. In essence, GPs can be used to define probability distributions over families of functions that fit a given set of data without explicitly assuming a functional basis. For an in-depth introduction to GP theory and applications, see \eg~\cite{Rasmussen:2006gpr}.

Recently, GPR has also been applied to the probabilistic inversion of the KL spectral representation~\cite{Horak:2021syv} as well as the extraction of parton distribution functions~\cite{Alexandrou:2020tqq, DelDebbio:2021whr, Candido:2023nnb}. The present work follows the same line of reasoning: making use of the fact that GPs are closed under linear transformations, it is possible to infer data from indirect observations that are related to the quantity of interest by a linear forward process~\cite{valentine2020gaussian}. In particular, we may obtain predictions for the spectral function from measurements of the associated correlator without inverting the KL transformation directly.

\begin{figure*}
	\centering
	\subfloat[]{%
		\includegraphics[width=0.5\linewidth]{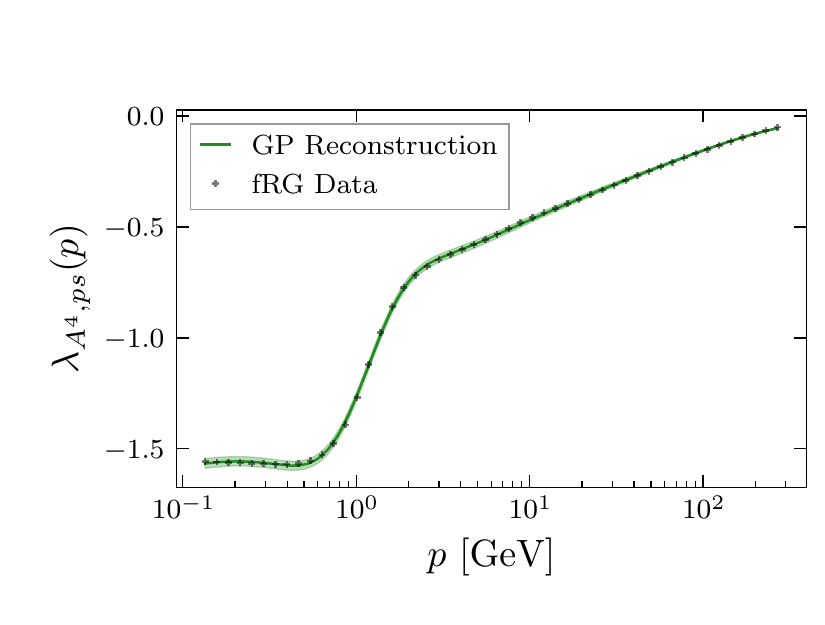}\label{fig:pseudoscalardressing}}%
	\subfloat[]{%
		\includegraphics[width=0.5\linewidth]{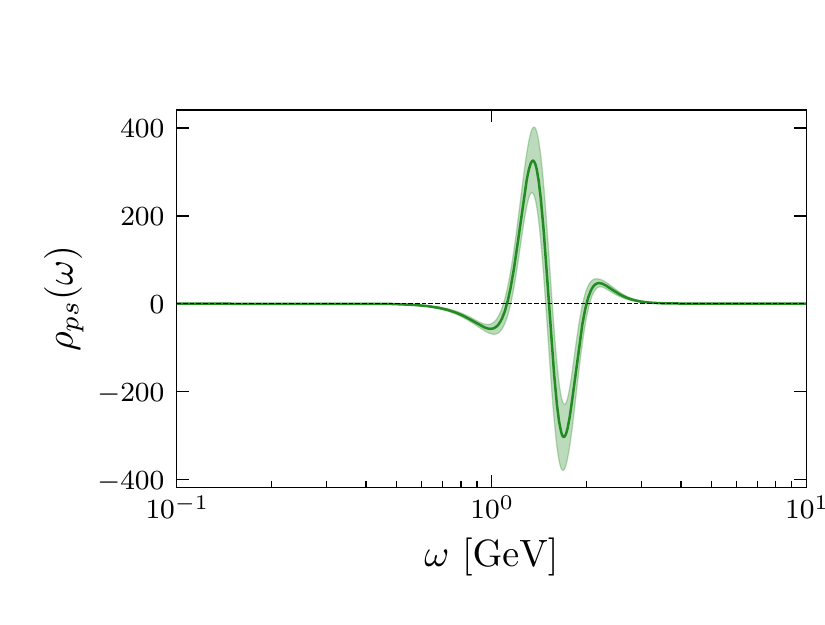}\label{fig:pseudoscalarsf}}
	\caption{(a) Euclidean dressing of the four-gluon vertex $\lambda_{A^4,ps}$ with the projection to obtain the pseudo-scalar glueball mass, see \Cref{sec:4gluon}, from the fRG (black crosses). This is compared to the reconstruction from the GP (green line). The corresponding spectral function $\rho_{A^4,ps}$ over frequency $\omega$ obtained with GPR is shown in (b). The light green band represents the $1\sigma$ region.}
	\label{fig:pseudoscalarplot}
\end{figure*}

To this end, we start by defining a GP prior distribution over spectral functions that encodes our knowledge and assumptions about $\rho(\omega)$ before making any observations,
\begin{equation}
  \rho(\omega) \sim \mathcal{GP}\left(\mu(\omega), k(\omega, \omega) \right) \,.
\end{equation}
Evaluating this GP for any set of points $\omega_i$ results in a multivariate normal distribution with mean $\mu(\omega_i)$ and covariance $k(\omega_i, \omega_j)$. Furthermore, the discrete propagator data $G(p_i)$ are then also normally distributed, with mean and covariance obtained by applying the linear forward process (the KL integral) to $\mu$ and $k$, \ie
\begin{equation}
\begin{aligned}
	G(p_i) \sim &\,\mathcal{N}\left(\int\text{d}\omega K(p_i, \omega) \mu(\omega)\,,\right.\\[1ex]
	& \,\left.\int\text{d}\omega\,\text{d}\omega' \,K(p_i, \omega)  K(p_i, \omega') k(\omega,\omega') \right) \,.
\end{aligned}
\end{equation}
As GPs can be specified completely by their second-order statistics, $\mu$ is usually set to zero for simplicity since any information contained therein may be fully absorbed into $k$. However, a non-zero prior mean may still be useful in practice, in which case it can simply be subtracted from the data beforehand. Using bold symbols for vectors of discrete data, \eg~$\boldsymbol{p}$ for a set of momenta $p_i$, the joint distribution of spectral function values $\rho$ at any point $\omega$ and a set of correlator data $G(\boldsymbol{p})$ can be expressed as
\begin{equation}\label{eq:GPprior}
\begin{aligned}
    \begin{bmatrix}
      \rho(\omega) \\[1ex]
      G(\boldsymbol{p})
    \end{bmatrix}
    \sim \mathcal{N}\left(0,
    \begin{bmatrix}
      k(\omega,\omega) & \boldsymbol{w}^\top(\omega)\\[1ex]
      \boldsymbol{w}(\omega) & \boldsymbol{W} + \sigma_n^2 \mathbbm{1}\\[1ex]
    \end{bmatrix}
    \right) \,,
\end{aligned}
\end{equation}
with
\begin{equation}
\begin{aligned}
  &[\boldsymbol{w}]_i(\omega) = \int \text{d}\omega'\, K(p_i, \omega') k(\omega', \omega)\,,\\[1ex]
  &[\boldsymbol{W}]_{ij} = \int\text{d}\omega'\text{d}\omega'' \, K(p_i, \omega') K(p_j, \omega'') k(\omega', \omega'') \,.
\end{aligned}
\end{equation}
Since the joint distribution is normal, the posterior distribution of the spectral function conditioned on observations of the correlator can be derived in closed form,
\begin{equation}\label{eq:posteriorKL}
\begin{aligned}
    \rho(\omega)|G(\boldsymbol{p}) \sim &\, \mathcal{GP}\Bigl(\boldsymbol{w}^\top(\omega)(\boldsymbol{W} + \sigma_n^2 \mathbbm{1})^{-1}G(\boldsymbol{p}), \\[1ex]
      & \hspace{-.5cm}k(\omega,\omega) - \boldsymbol{w}^\top(\omega)(\boldsymbol{W} + \sigma_n^2 \mathbbm{1})^{-1}\boldsymbol{w}(\omega)\Bigr) \,.
\end{aligned}
\end{equation}
This is a standard result in multivariate statistics and is essentially equivalent to GPR with direct observations, only with additional insertions of the linear transformation that one seeks to invert. The GP posterior \labelcref{eq:posteriorKL} encodes our knowledge of the spectral function given the correlator data and directly accounts for some additive Gaussian noise with variance $\sigma_n^2$ in the observations.

For computational applications of GPs, the covariance is usually parameterised by a kernel function $k(\omega,\omega')$, as already implied by the notation used above. Since this fully specifies the GP as mentioned previously, choosing the right type of kernel is a pivotal part of finding a good model for a given set of data. A natural choice in many applications are so-called \textit{universal} kernels that can describe any continuous function~\cite{steinwart:2002} and hence provide the required flexibility when little is known about other properties of the desired solution a priori. The radial basis function (RBF) kernel,
\begin{equation}\label{eq:rbf}
	k(\omega, \omega') = \sigma^2 \exp\left(-\frac{(\omega - \omega')^2}{2\ell^2}\right) \,,
\end{equation}
is a popular choice due to its universality and every function in the associated prior being infinitely differentiable, and is also employed in the present work. The parameters $\ell$ and $\sigma$ control the length scale and overall magnitude of the correlation between data and are subject to optimisation; see \Cref{app:nll}.

Predictions obtained with GPR can also be understood within the well-known Backus-Gilbert framework~\cite{backusgilbert}, one of the most popular approaches to spectral reconstruction in the lattice community. In fact, both methods produce numerically equivalent estimates under certain conditions~\cite{valentine2020gaussian}, despite following different philosophies. Nevertheless, the GPR picture is much more flexible, since essentially any available prior information can be systematically incorporated into the regression by extending the covariance matrix in \labelcref{eq:GPprior}, following the same reasoning as in the construction of the joint distribution of $G$ and $\rho$. Such prior information may simply consist of known values of the spectral function at certain points, in which case they are treated as direct observations. More generally, it can be any indirect data related to $\rho$ through a linear operator---such as a derivative~\cite{solak:2003}---and even inequality constraints such as bounds and monotonicity conditions~\cite{Agrell:2019}.

In summary, GPR is a powerful approach to tackle ill-conditioned linear inverse problems probabilistically, which makes it an attractive candidate algorithm for spectral reconstruction in quantum field theory.

\section{Results}\label{sec:Results}

We calculate the vertex dressings with the fRG as outlined in \Cref{sec:4gluon}; for details on the truncation and computation, see \Cref{app:NumDetails}. The resulting Euclidean dressing functions for the scalar and pseudo-scalar projections are shown in \Cref{fig:scalardressing,fig:pseudoscalardressing}, respectively.

In the channels considered here, the ghost loops drop out; see \Cref{app:Proj}. Hence, these channels are free of the IR divergences that are in general present in the four-gluon vertex and we can utilise the constraint $\rho_{A^4}(0) = 0$ in the GP reconstruction. Furthermore, an additional bias is introduced in order to suppress unphysical oscillations at the tails of the spectral function. Similar to the procedure applied in \cite{Horak:2021syv}, this is achieved by rescaling the frequency with a soft step function,
\begin{equation}\label{eq:GBbias}
	\omega \rightarrow \tilde{\omega} = \frac{1}{\exp(-2(\omega - \omega_0)/\ell_0) + 1} \,,
\end{equation}
where the parameter $\ell_0$ controls the steepness and $\omega_0$ the position of the midpoint. This rescaling can be understood as the introduction of a frequency-dependent length scale in the RBF kernel, with smaller values around $\omega_0$ and larger values at the tails of the spectral function. We note that the resonances of interest are already observed without introducing this additional bias. However, the peaks are enhanced by this procedure while the reconstruction of the correlator remains in good agreement with the input data. While this parameterisation suppresses additional structures (such as excited glueball states at higher energies; see \eg \cite{Huber:2020ngt}), even without the rescaling \labelcref{eq:GBbias} no additional features beyond the dominant peak corresponding to the bound state are observed, apart from the usual oscillatory behaviour at the tail of the spectral function. This implies that higher excited states exhibit at most sub-leading contributions to these vertex projections. Resolving these structures therefore requires either more sophisticated projections of the tensor structures or a significantly higher precision in the calculation of the vertex itself.

The parameters of the RBF kernel and frequency rescaling are optimised by minimising an objective function, conventionally taken to be the negative log-likelihood (NLL). Unsurprisingly, the NLL shows a flat direction where some parameters are unconstrained; see \Cref{fig:grid_scans}. This can be interpreted as a manifestation of the ill-conditioned nature of the inverse problem, and may be treated by imposing a hyperprior. We observe that changing the parameters in this direction has negligible impact on the resonant peak position; see \Cref{fig:sf_nll_sc,fig:sf_nll_ps}. Hence, the seemingly heuristic use of a generic hyperprior is well justified in this context as it does not introduce a bias for the quantity of interest. Details about this procedure as well as the optimised parameter values are provided in \Cref{app:nll}. The intrinsic error estimate of the GP posterior is fixed to $\sigma_n = 10^{-2}$, corresponding to an upper bound on the uncertainty of the fRG calculation. $\sigma_n$ is not optimised as this diminishes the significance of the likelihood for the other parameters~\cite{ober2021promises}.

The reconstructed dressings are compared to the fRG input data in \Cref{fig:scalardressing,fig:pseudoscalardressing}, with the associated spectral functions shown in \Cref{fig:scalarsf,fig:pseudoscalarsf}, respectively. From a Hubbard–Stratonovich transformation it can be inferred that the vertex dressing corresponds to the negative dressing function of the bound state under consideration. Hence, the spectral functions $\rho_{sc/ps}$ are computed from the negative vertex dressing. Consequently, the positive peak indicates an asymptotic state that is interpreted as the respective glueball resonance. We also observe negative structures in the spectral function since the four-gluon vertex itself is not a gauge-invariant object. The reconstruction of the vertex dressing largely reproduces the fRG data within errors. For high momenta, the result deviates more strongly, in particular for the scalar glueball. This is due to the additional bias introduced to the kernel that specifically suppresses any dynamics in the UV regime.

The glueball masses are extracted from the dominant peak positions of the spectral functions. We obtain $\hat{\omega}_{sc} = \unit[0.93]{GeV}$ for the scalar and $\hat{\omega}_{ps} = \unit[1.35]{GeV}$ for the pseudo-scalar channel. Since we work within the s-channel approximation and have two incoming momenta each with the magnitude $p$, the peak position corresponds to the half of the glueball mass, \ie~\mbox{$m_{sc/ps} = 2\,\hat{\omega}_{sc/ps}$}. Hence, we obtain the masses \mbox{$m_{sc} = \unit[1870(75)]{MeV}$} for the scalar and \mbox{$m_{ps} = \unit[2700(120)]{MeV}$} for the pseudo-scalar glueball. The reported errors are a combination of the standard deviations computed from the GP posterior and an additional $3\%$ error from the scale setting procedure of the input data. A more in-depth discussion of the systematic error of the reconstruction can be found in \Cref{app:nll}. We compare our results with masses obtained from independent lattice and DSE/BSE studies of the glueball spectrum in \Cref{tab:GB:Results} and find them to be in reasonable agreement, in particular for the pseudo-scalar channel where they match well within the provided uncertainties.

\setlength{\tabcolsep}{6pt}
\begin{table}[t]
	\centering
	\begin{tabular}{l l@{ }l l@{ }l l}
		\toprule 
		$J^{PC}$ & lattice & & DSE-BSE & & this work \\[1ex]
		\midrule\midrule
		\multirow{4}{*}{$0^{++}$} & $1760(70)$ & \cite{Morningstar:1999rf}  & $1850(130)$ & \cite{Huber:2020ngt} & $1870(75)$ \\[1ex]
		& $1740(70)$ & \cite{Chen:2005mg}  & $1640$ & \cite{Sanchis-Alepuz:2015hma}& \\[1ex]
		& $1651(23)$ & \cite{Athenodorou:2020ani} & & & \\[1ex]
	    & $1618(26)(25)$ & \cite{Sakai:2022zdc}  & & & \\[1ex]
		\midrule
		\multirow{4}{*}{$0^{-+}$} & $2650(60)$ & \cite{Morningstar:1999rf} & $2580(180)$ & \cite{Huber:2020ngt} & $2700(120)$ \\[1ex]
		& $2610(70)$ & \cite{Chen:2005mg}  & $4530$ & \cite{Sanchis-Alepuz:2015hma}& \\[1ex]
		& $2600(40)$ & \cite{Athenodorou:2020ani} & & & \\[1ex]
	    & $2483(61)(55)$ & \cite{Sakai:2022zdc}  & & & \\[1ex]
		\bottomrule
	\end{tabular}
	\caption{Comparison of scalar ($J^{PC}= 0^{++}$) and pseudo-scalar ($J^{PC}= 0^{-+}$) glueball masses from different methods. The results of \cite{Morningstar:1999rf, Chen:2005mg} are rescaled to match \cite{Athenodorou:2020ani, Sakai:2022zdc} with $r_0 = 1/\unit[418(5)]{MeV}$. The errors of \cite{Morningstar:1999rf, Chen:2005mg} are a combination of statistical as well as systematic uncertainties stemming from the lattice anisotropy and the scale $r_0$. The errors for \cite{Athenodorou:2020ani} are statistical only. For \cite{Sakai:2022zdc}, the quoted values are the statistical as well as systematic uncertainties for the continuum extrapolation, respectively. For \cite{Huber:2020ngt}, the error comes from the extrapolation method.}
	\label{tab:GB:Results}
\end{table}

\section{Conclusion}\label{sec:Conclusion}

We put forward a self-consistent approach for the extraction of bound state information from gauge-fixed correlation functions. Key to this framework is the spectral reconstruction of interaction channels in Euclidean space-time that have overlap with the corresponding gauge-invariant bound state. The method is applied to low-lying glueball states in Yang-Mills theory, extracted from the dressing functions of the Euclidean four-gluon vertex. With appropriate projection operators of the four-gluon vertex, we obtain access to the masses of the scalar and pseudo-scalar glueballs.

The Euclidean dressings are obtained with the functional renormalisation group, also utilising earlier results for correlation functions from \cite{Pawlowski:2022oyq}. The respective spectral functions are then computed via Gaussian process regression and their resonance peaks are identified with the glueball masses: for the scalar and pseudo-scalar glueballs, we arrive at $\unit[1870(75)]{MeV}$ and $\unit[2700(120)]{MeV}$, respectively. The results agree well with independent studies of the glueball spectrum, lending further credibility to our proposed method of computing bound state properties from vertex dressing functions via spectral reconstruction. The present approach can also be directly applied to higher glueball states in Yang-Mills theory, as well as glueball and other hadronic states in QCD.

\section*{Acknowledgments}

We thank Jan~Horak, Markus~Q.~Huber, and William~I.~Jay for discussions. This work is funded by the Deutsche Forschungsgemeinschaft (DFG, German Research Foundation) under Germany’s Excellence Strategy EXC 2181/1 - 390900948 (the Heidelberg STRUCTURES Excellence Cluster) and the Collaborative Research Centre SFB 1225 (ISOQUANT). JT and NW acknowledge support by the Deutsche Forschungsgemeinschaft (DFG, German Research Foundation) – Project number 315477589 – TRR 211. NW acknowledges the support by the State of Hesse within the Research Cluster ELEMENTS (Project ID 500/10.006). JMU is supported in part by Simons Foundation grant 994314 (Simons Collaboration on Confinement and QCD Strings) and the U.S.\ Department of Energy, Office of Science, Office of Nuclear Physics, under grant Contract Number DE-SC0011090. This work is funded by the U.S.\ National Science Foundation under Cooperative Agreement PHY-2019786 (The NSF AI Institute for Artificial Intelligence and Fundamental Interactions, \url{http://iaifi.org/}).

\appendix

\section{Details of the fRG setup}\label{app:NumDetails}

\subsection{fRG equation for the four-gluon vertex}

The master equation of the fRG is the flow equation of the scale-dependent 1PI effective action. It is obtained by introducing an IR cutoff with a cutoff scale $k$ via a momentum-dependent mass function $R_k(p^2)$ that is added to the inverse propagator. The respective flow equation is derived via taking a derivative of the generating functions w.r.t.~the cutoff scale $k$, 
\begin{equation}\label{eq:frgflow}
	\partial_t \Gamma_k [\Phi] = \frac{1}{2} \text{Tr}\frac{1}{\Gamma_k^{(2)}[\Phi] + R_k}\partial_t R_k, ~~ t = \log\left(\frac{k}{\Lambda}\right) \,,
\end{equation}
where $t$ is the RG-time, and the trace in \labelcref{eq:frgflow} sums over species of fields, space-time (momentum), Lorentz indices and group indices. The regulator functions carry the classical dispersion of the ghost and gluon fields as well as a dimensionless shape function. The present results are computed with the usual exponential shape function,
\begin{equation}
	r(p^2/k^2) = \frac{\text{e}^{-p^2/k^2}}{1-\text{e}^{-p^2/k^2}} \,,
\end{equation}
and an additional wave function renormalisation $Z_{A,k}$ or $Z_{c,k}$; for more details see \cite{Pawlowski:2022oyq}. For a recent review of the fRG see \eg~\cite{Dupuis:2020fhh} and references therein.

Our general setup in Landau gauge Yang-Mills theory follows \cite{Cyrol:2016tym, Pawlowski:2022oyq}. The flow of the four-point vertex is obtained by taking the fourth derivative of \labelcref{eq:frgflow} w.r.t.~the gluon field. In this work, we are only interested in certain channels of the four-gluon vertex. Hence, we do not solve the full system self-consistently, but take all other correlation functions such as the gluon propagators from \cite{Pawlowski:2022oyq} as input.

The fRG equation for the four-gluon vertex solved in the present work is depicted in \Cref{fig:fRGGlueballs}. This flow is integrated on the solution of the correlation functions obtained in \cite{Pawlowski:2022oyq}. There, different IR closures of correlation functions in the Landau gauge have been computed, and the present work utilises the \textit{scaling} solution. The independence of this choice has recently been shown in~\cite{Huber:2020ngt}, where both solutions---decoupling and scaling---were considered in the context of glueballs.
The approximation used in \cite{Pawlowski:2022oyq} only includes the primitively divergent (classical) tensor structures, which leads to semi-quantitative results. Further details can be found in \cite{Pawlowski:2022oyq}. 

We use the $k$-dependent dressing functions from \cite{Pawlowski:2022oyq} as input. Their parameterisations are given by
\begin{equation}\label{eq:4vertex}
\begin{aligned}
	&\Gamma_{AA, \mu\nu}^{(2),ab}(p)  = \delta^{ab} \Pi^\perp_{\mu\nu}(p) Z_A (p) (p^2 + m_T^2), \\[1ex]
	&\Gamma_{A^3, \mu\nu\rho}^{(3),abc}(p_1, p_2)  = i f^{abc} \lambda_{A^3} (\bar{p}) \Bigl[(p_1 - p_2)_\rho \delta_{\mu\nu} + \mathrm{perm.}\Bigr], \\[1ex]
	&\Gamma_{A^4, \mu\nu\rho\sigma}^{(4),abcd}(p_1,p_2,p_3) = \lambda_{A^4}(\bar{p}) \Bigl[f^{abn}f^{cdn}\delta_{\mu\rho}\delta_{\nu\sigma} + \mathrm{perm.} \Bigl] \,,
\end{aligned}
\end{equation}
where we approximate the full momentum dependence of the vertices with the symmetric point configuration $\bar{p}$, see \eg~\cite{Cyrol:2016tym}, defined by
\begin{equation}
    \bar{p}^2 = \frac{1}{n}\sum_{i=1}^{n}p_i^2 \,,
\end{equation}
with $n=3,4$.

\begin{figure*}[t]
    \centering
    \subfloat[RBF parameters, scalar channel.]{%
    	\includegraphics[width=0.5\textwidth]{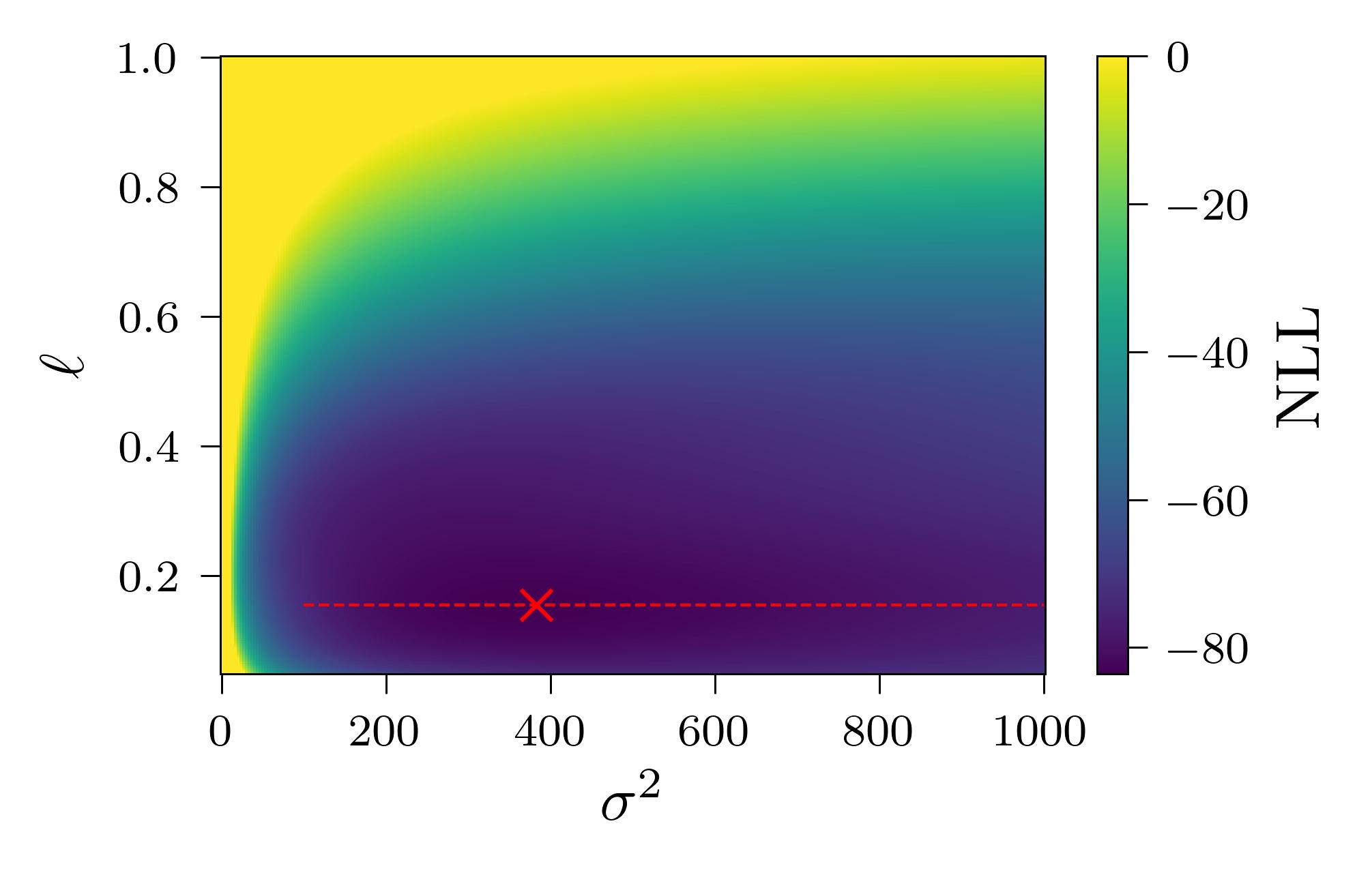}\label{fig:gridscan_rbf}}%
    \subfloat[Bias parameters, scalar channel.]{%
    	\includegraphics[width=0.5\textwidth]{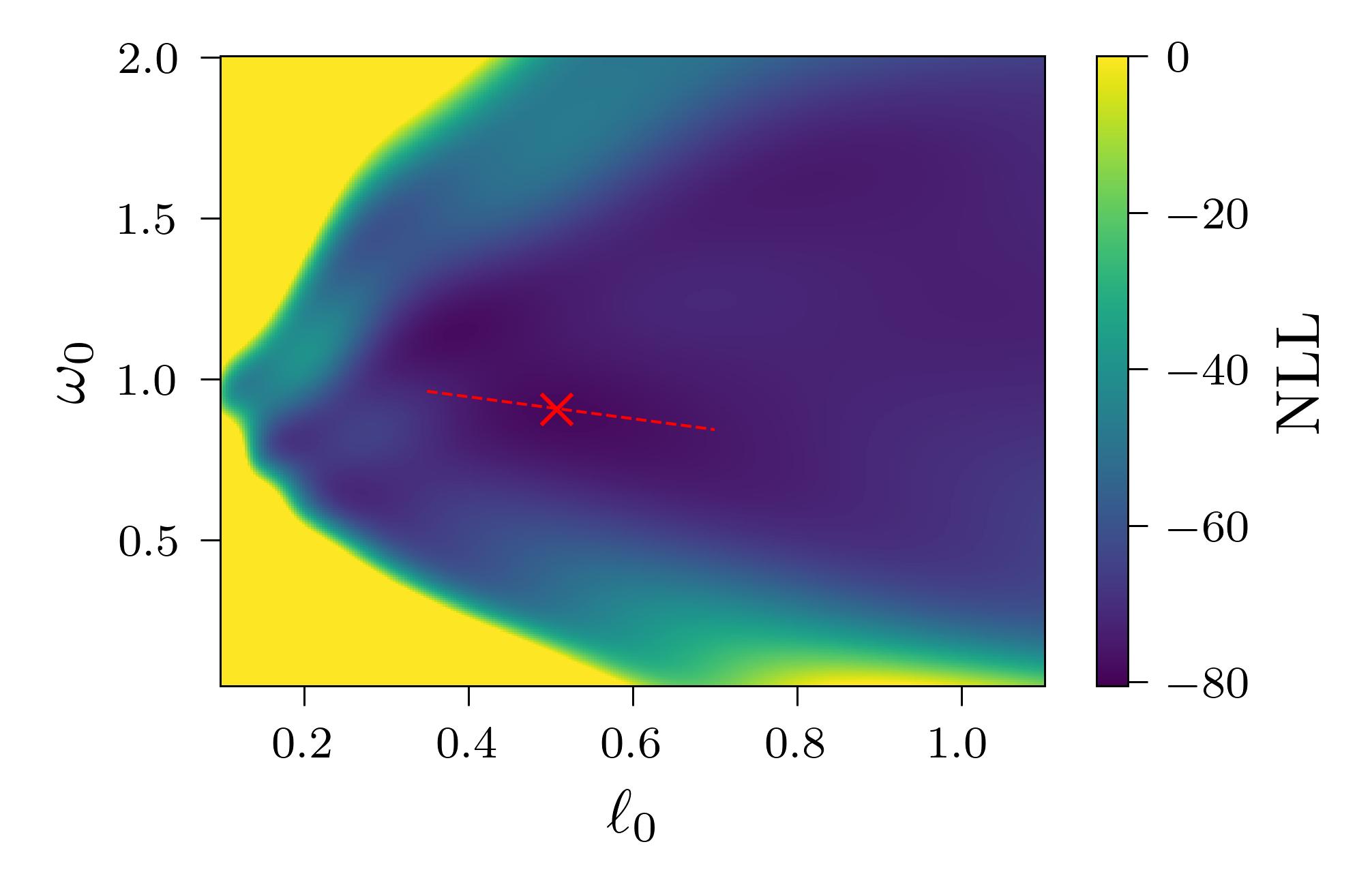}\label{fig:gridscan_bias}}
    	
    \subfloat[RBF parameters, pseudo-scalar channel.]{%
    	\includegraphics[width=0.5\textwidth]{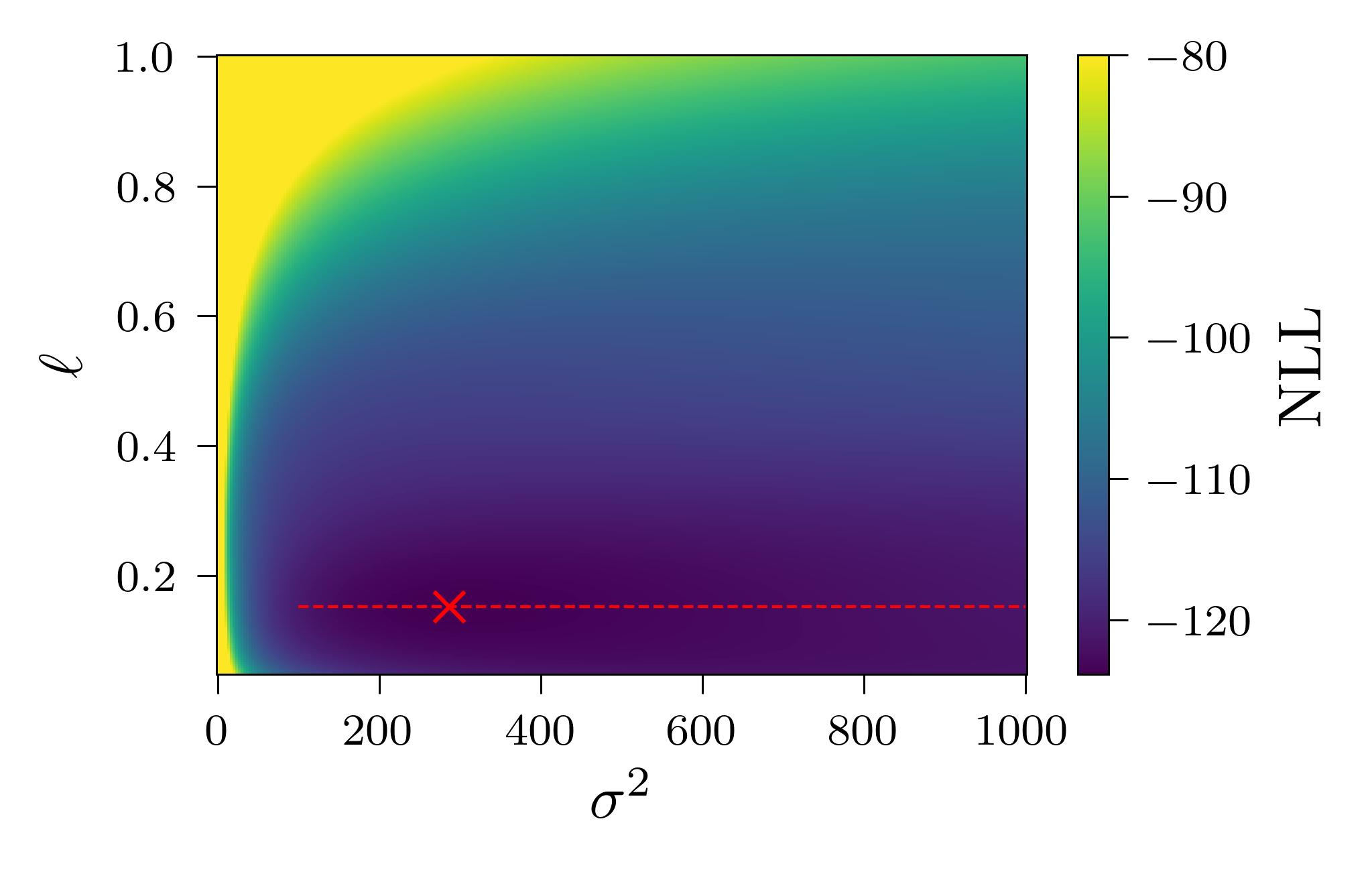}\label{fig:ps_gridscan_rbf}}%
    \subfloat[Bias parameters, pseudo-scalar channel.]{%
    	\includegraphics[width=0.48\textwidth]{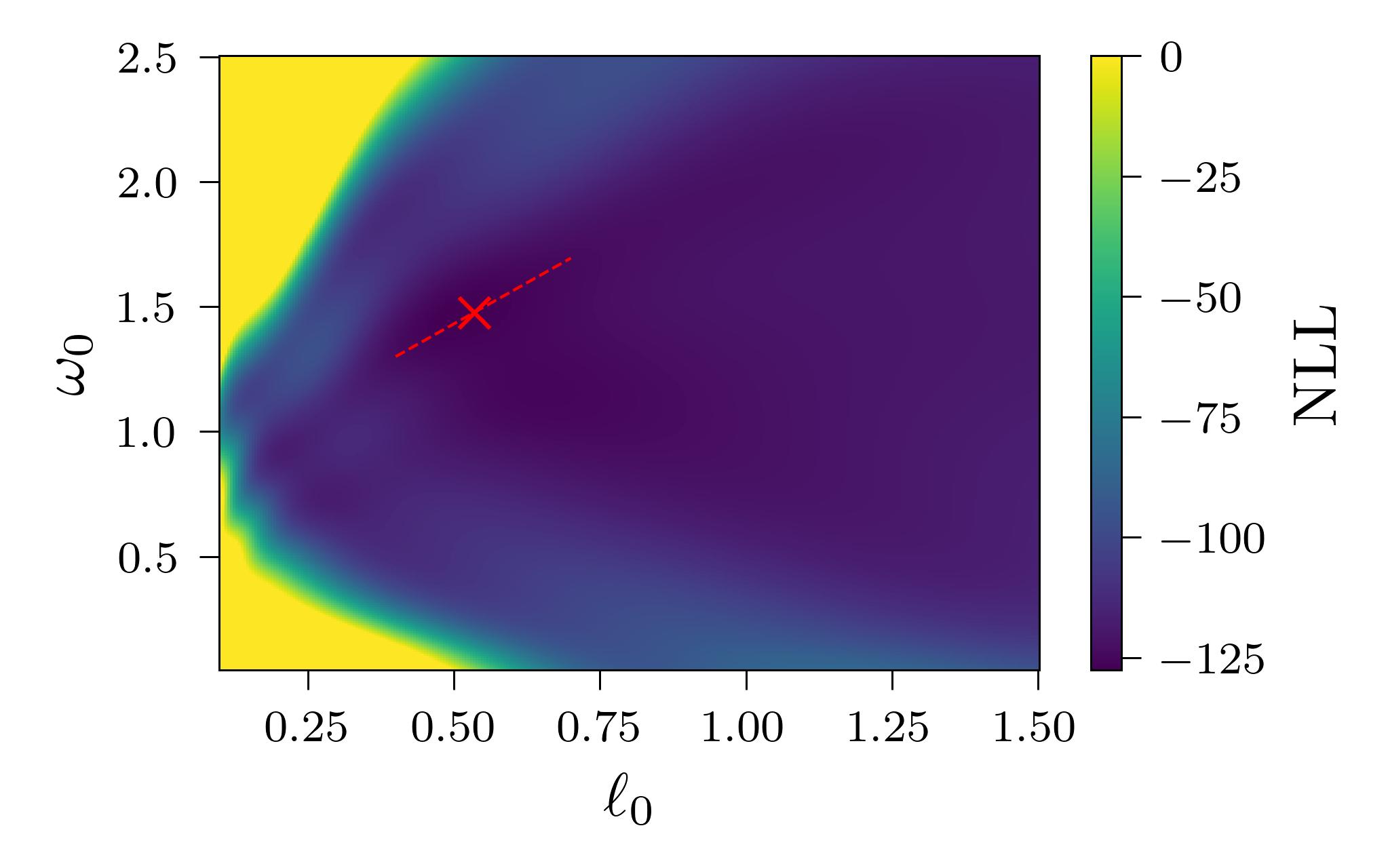}\label{fig:ps_gridscan_bias}}
    \caption{Grid scans of the NLL \labelcref{eq:NLL} of the reconstructions for both channels. Note that the optimisations are performed subsequently, starting with the RBF and followed by the bias parameters. The red lines indicate the trajectories in parameter space used for comparing the variance in the spectral functions; see \Cref{fig:sf_nll_sc,fig:sf_nll_ps}. The red cross indicates the NLL optimised parameters; see \Cref{tab:GB:OptParam}.}
    \label{fig:grid_scans}
\end{figure*}

\begin{figure*}[t]
\centering
	\subfloat[Scan of the flat direction in the RBF parameters.]{%
    	\includegraphics[width=0.5\textwidth]{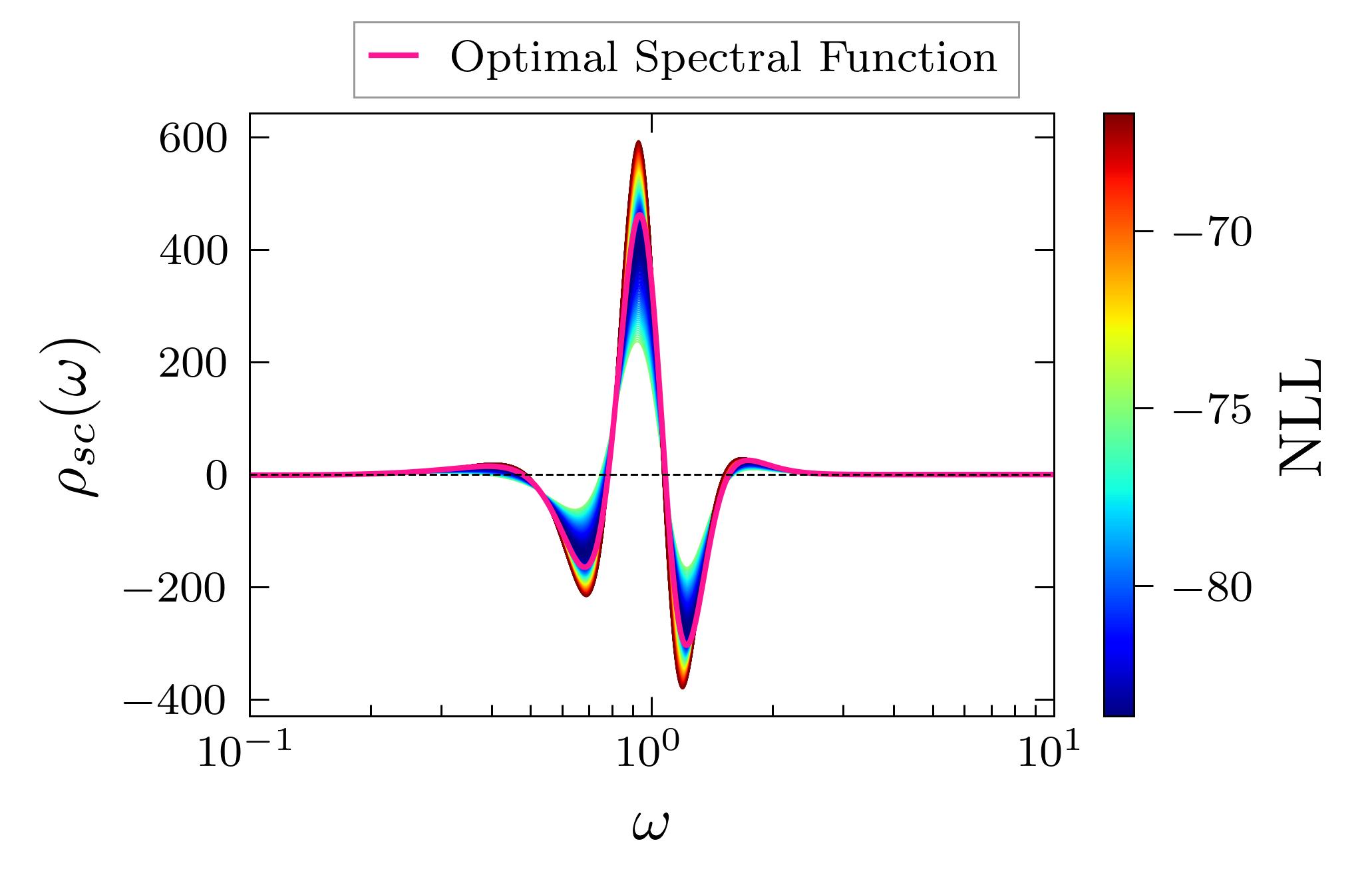}}%
	\subfloat[Scan around optimal bias parameters.]{%
	    \includegraphics[width=0.5\textwidth]{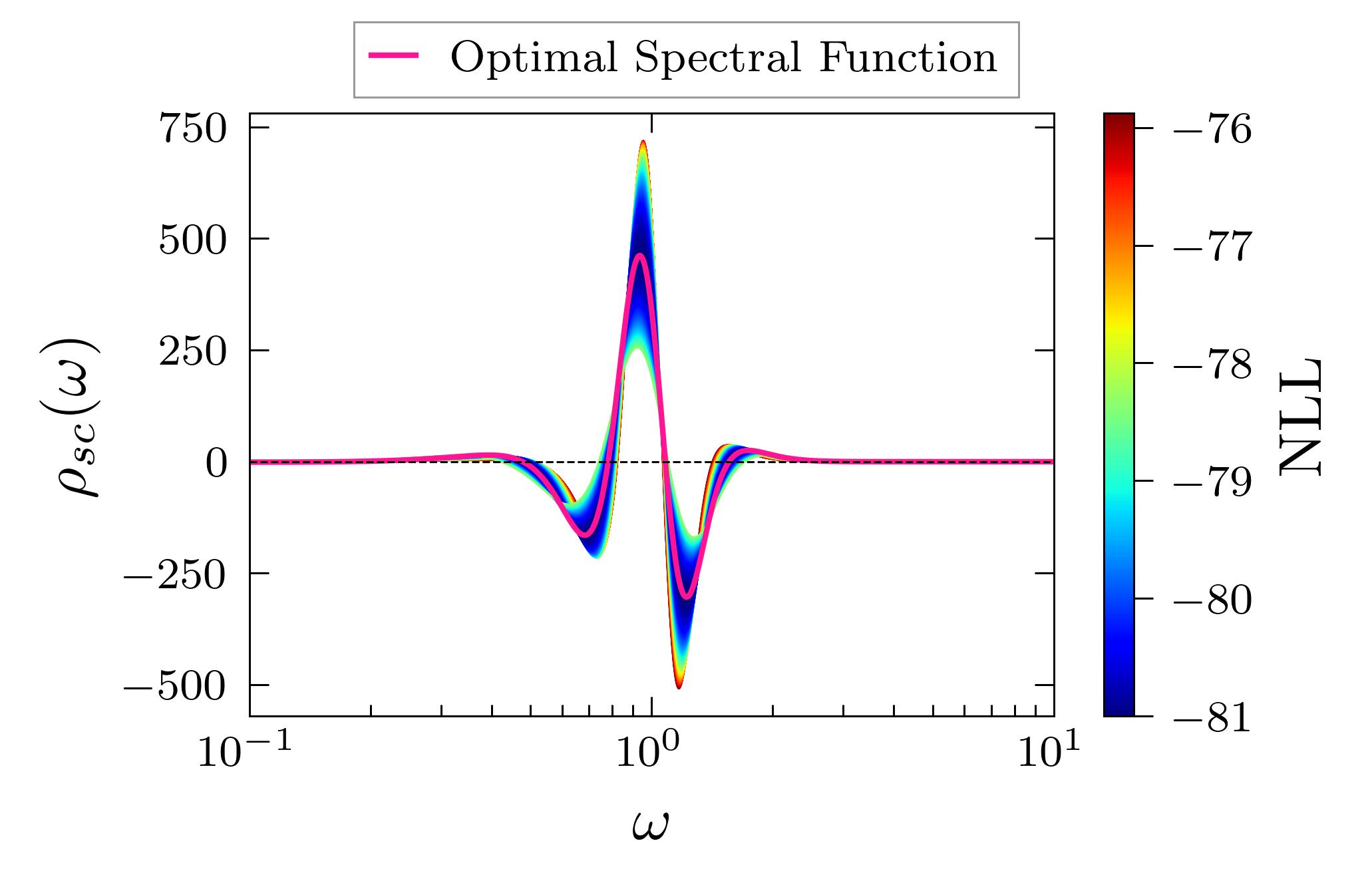}}
	\caption{Spectral function of the scalar channel. The bands show the variance with respect to the flat directions of the parameter space; see~\Cref{fig:gridscan_rbf,fig:gridscan_bias}. The peak position is observed to be robust, even under large parameter changes. The overall magnitude on the other hand shows considerable variation for both sets of parameters.}
	\label{fig:sf_nll_sc}
\end{figure*}

\begin{figure*}[t]
\centering
	\subfloat[Scan of the flat direction in the RBF parameters.]{%
	    \includegraphics[width=0.5\textwidth]{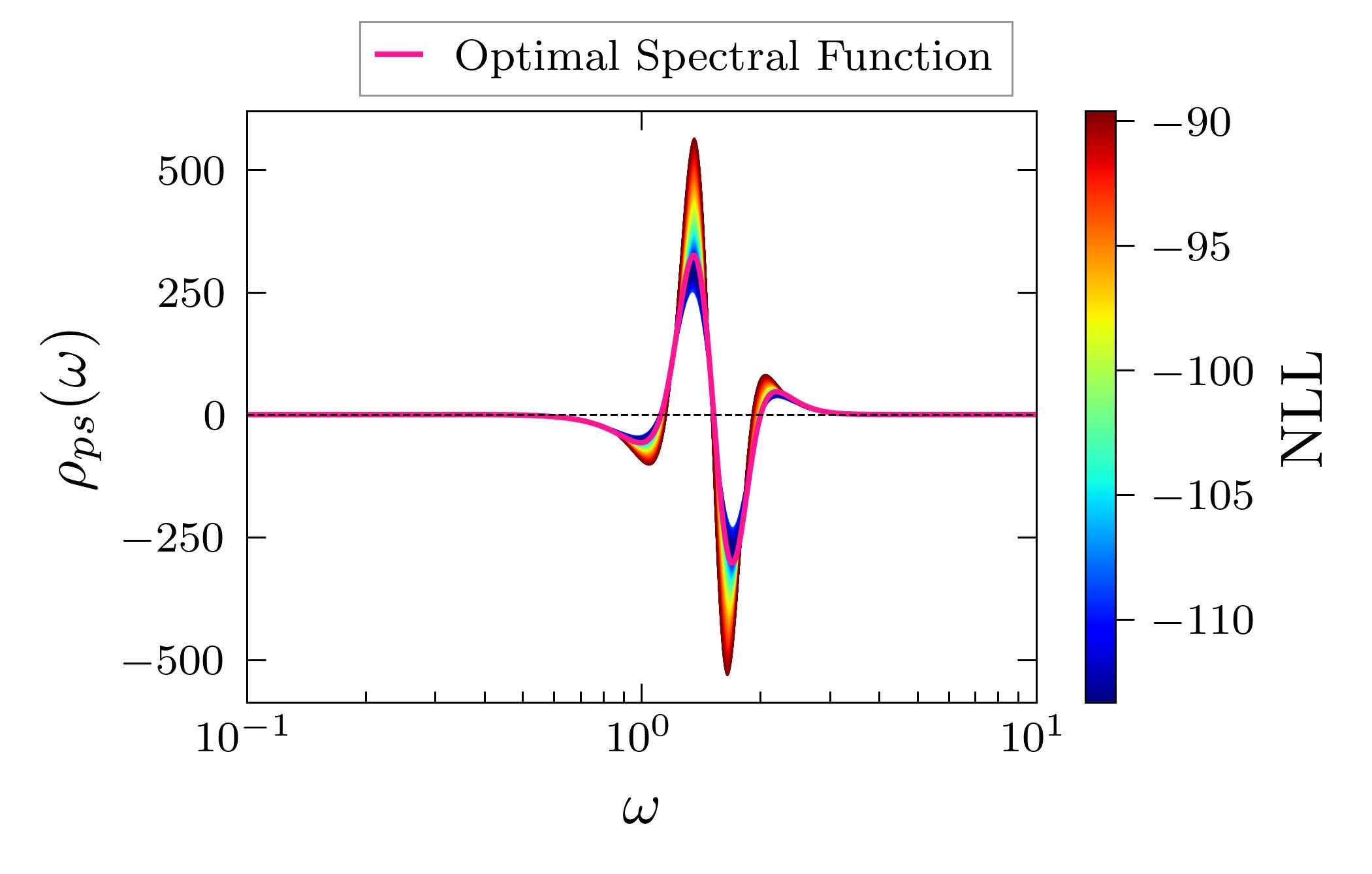}}%
	\subfloat[Scan around optimal bias parameters.]{%
    	\includegraphics[width=0.5\textwidth]{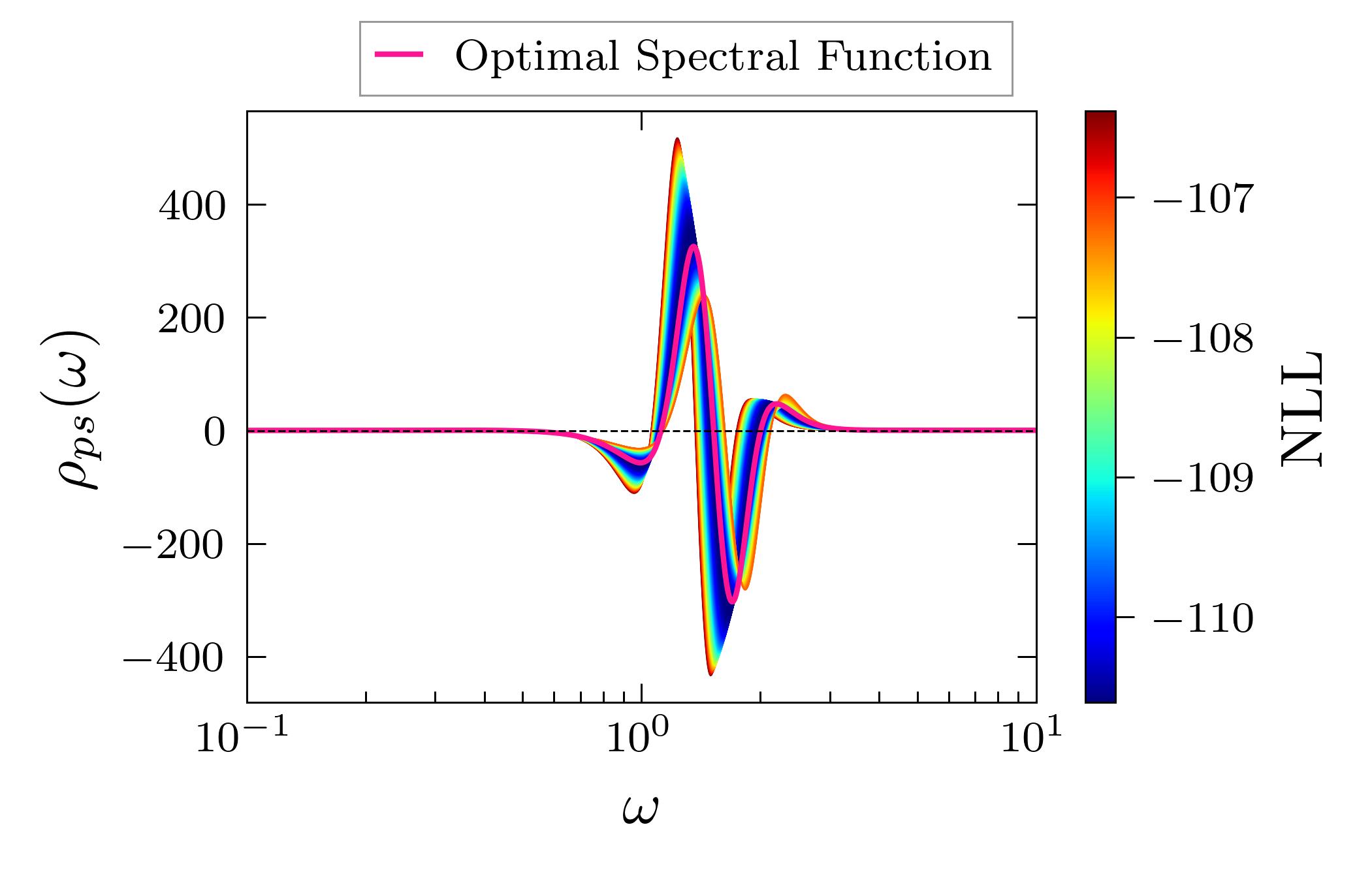}}
	\caption{Spectral function of the pseudo-scalar channel. The bands show the variance with respect to the flat directions of the parameter space; see~\Cref{fig:ps_gridscan_rbf,fig:ps_gridscan_bias}. The peak position is observed to be robust under variations of the RBF parameters, but shows significantly stronger deviations compared to the scalar channel for different values of the bias parameters.}
	\label{fig:sf_nll_ps}
\end{figure*}

\subsection{Glueball projection operators}\label{app:Proj}

The full projection operator for obtaining the scalar glueball mass is simply a contraction with the transvere part of the classical tensor structure, given by
\begin{equation}\label{eq:sctensor}
\begin{aligned}
	&\mathcal{P}^{abcd}_{s,\mu\nu\rho\sigma}(p,p,-p,-p) = \\[1ex] &\frac{\left[\Pi^\perp(p)\Pi^\perp(p)\Pi^\perp(-p)\Pi^\perp(-p)\tau_{A^4,cl}(p,p,-p)\right]^{abcd}_{\mu\nu\rho\sigma}}{\Pi^\perp(p)\Pi^\perp(p)\Pi^\perp(-p)\Pi^\perp(-p)\tau_{A^4,cl}(p,p,-p)\tau_{A^4,cl}(p,p,-p)} \,.
\end{aligned}
\end{equation}
Indices are suppressed for simplicity and the external momenta are already matched to the momentum parameterisation of the four-gluon vertex,
\begin{equation}
	p_1 = p_2 = -p_3 = -p_4 \equiv p \,.
\end{equation}
The classical four-gluon tensor structure $\tau_{A^4,cl}$ is given in \labelcref{eq:classical_tensor} and the transverse projection operator is
\begin{equation}
	\Pi^\perp_{\mu\nu}(p) = \delta_{\mu\nu} - \frac{p_\mu p_\nu}{p^2} \,.
\end{equation}
The pseudo-scalar projection operator is defined by the tensor structure \labelcref{eq:pseudotensor} and given by
\begin{widetext}
\begin{equation}\label{eq:pstensor}
\begin{aligned}
	\mathcal{P}^{abcd}_{ps,\mu\nu\rho\sigma}(p_1,p_2,-p_1,-p_2) =\frac{\left[\Pi^\perp(p_1)\Pi^\perp(p_2)\Pi^\perp(-p_1)\Pi^\perp(-p_2)\tau_{A^4,ps}(p_1,p_2)\right]^{abcd}_{\mu\nu\rho\sigma}}{\Pi^\perp(p_1)\Pi^\perp(p_2)\Pi^\perp(-p_1)\Pi^\perp(-p_2)\tau_{A^4,ps}(p_1,p_2)\tau_{A^4,ps}(p_1,p_2)} \,,
\end{aligned}
\end{equation}
\end{widetext}
where the in- and outgoing external momenta are chosen to be orthogonal,
\begin{equation}
\begin{aligned}
	p_1 = - p_3, ~~p_2 = -p_4, \\
	p_1 \cdot p_2 = 0,~~	p_1^2 = p_2^2 = p^2 \,.
\end{aligned}
\end{equation}
We note that the projection onto the ghost loop part of the flow analytically vanishes for both projections on the momentum configurations under consideration. This was observed for a similar momentum configuration in \cite{Cyrol:2014kca}. 

\section{Optimisation of GP kernel parameters}\label{app:nll}

As stated in \Cref{sec:Results}, the kernel hyperparameters of the GP \labelcref{eq:GPprior} are optimised by minimising the associated NLL,
\begin{equation}\label{eq:NLL}
\begin{aligned}
	-\log p(G(\boldsymbol{p})|\boldsymbol{\sigma}) = \frac{1}{2} G(\boldsymbol{p})^\top \left(\boldsymbol{W_\sigma} + \sigma_n^2 \mathbbm{1} \right)^{-1} G(\boldsymbol{p}) \\
	+ \frac{1}{2}\log \det(\boldsymbol{W_\sigma} + \sigma_n^2 \mathbbm{1}) + \frac{N}{2} \log 2\pi \,,
\end{aligned}
\end{equation}
where the dependence on the parameters $\boldsymbol{\sigma}$ is emphasised by an index. When optimising the parameters of the RBF kernel \labelcref{eq:rbf} and the frequency rescaling bias \labelcref{eq:GBbias} simultaneously, the parameters favour values that tend to nullify the bias, such as $\ell_0$ becoming large. Hence, the parameters are first optimised only considering the bare RBF kernel in order to obtain baseline values. Subsequently, the bias is introduced and its parameters are optimised given the RBF kernel calculated beforehand. This way, the position and size of the dynamical part of the spectral function are also subject to optimisation.

The parameters are optimised by performing a high-resolution grid scan; see \Cref{fig:grid_scans}. Their optimal values are provided in \Cref{tab:GB:OptParam}. In the direction of the magnitude parameter $\sigma$, the NLL does not change significantly for larger values. Similarly, in~\cite{Horak:2021syv} this parameter was observed to exhibit an open direction towards infinity and a hyperprior had to be introduced, which is a manifestation of the ill-conditioning of the inversion. The dependence of the spectral function on $\sigma$ is plotted in \Cref{fig:sf_nll_sc,fig:sf_nll_ps}, showing that while it does impact the magnitude of the dominating peak, its position and other general features of the spectral function remain stable. Accordingly, the overall magnitude of the computed spectral functions should be taken with a grain of salt, but predictions of other features such as the peak position, width, and overall shape are robust as the NLL diverges quickly when considering non-optimal parameters. Scanning the spectral functions in the plane of the bias parameters on the other hand reveals a more drastic change in the peak position. However, these parameters are restricted to a much smaller region by the likelihood and the stability of the peak position is retained. For a more quantitative statement about the systematic error of the reconstruction, an empirical comparison of different bias parameterisations is required. This can potentially be achieved by mapping out the posterior probability landscape with Monte Carlo methods.

\begin{table}[b]
	\centering
	\begin{tabular}{c c c c c}
		\toprule
		\rule{0pt}{12pt}$J^{PC}$ & $\sigma$ & $\ell$ & $\omega_0$ & $\ell_0$ \\
		\midrule\midrule
		\rule{0pt}{12pt}$0^{++}$ & $19.56$ & $0.155$ & $0.908$ & $0.508$ \\
		\midrule
		\rule{0pt}{12pt}$0^{-+}$ & $16.93$ & $0.152$ & $1.476$ & $0.534$ \\
		\bottomrule
	\end{tabular}
	\caption{Optimised GP hyperparameters for the kernel and rescaling functions, \labelcref{eq:rbf} and \labelcref{eq:GBbias}.}
	\label{tab:GB:OptParam}
\end{table}

\section{Implementation details}

The fRG equations are derived using \texttt{QMeS}~\cite{Pawlowski:2021tkk}, a Mathematica package for the derivation of symbolic functional equations. After projecting onto the respective glueball channels, the equations are traced with \texttt{FormTracer}~\cite{Cyrol:2016zqb}. The resulting momentum-dependent integral-differential equations are solved in Mathematica 12.0. The GPR is implemented in Python 3 employing the \texttt{NumPy}~\cite{harris2020array} and \texttt{SciPy} libraries~\cite{2020SciPy-NMeth}. Integrals are approximated using a discrete Riemann sum.

\bibliography{bib_master}

\begin{thebibliography}{47}%
\makeatletter
\providecommand \@ifxundefined [1]{%
 \@ifx{#1\undefined}
}%
\providecommand \@ifnum [1]{%
 \ifnum #1\expandafter \@firstoftwo
 \else \expandafter \@secondoftwo
 \fi
}%
\providecommand \@ifx [1]{%
 \ifx #1\expandafter \@firstoftwo
 \else \expandafter \@secondoftwo
 \fi
}%
\providecommand \natexlab [1]{#1}%
\providecommand \enquote  [1]{``#1''}%
\providecommand \bibnamefont  [1]{#1}%
\providecommand \bibfnamefont [1]{#1}%
\providecommand \citenamefont [1]{#1}%
\providecommand \href@noop [0]{\@secondoftwo}%
\providecommand \href [0]{\begingroup \@sanitize@url \@href}%
\providecommand \@href[1]{\@@startlink{#1}\@@href}%
\providecommand \@@href[1]{\endgroup#1\@@endlink}%
\providecommand \@sanitize@url [0]{\catcode `\\12\catcode `\$12\catcode
  `\&12\catcode `\#12\catcode `\^12\catcode `\_12\catcode `\%12\relax}%
\providecommand \@@startlink[1]{}%
\providecommand \@@endlink[0]{}%
\providecommand \url  [0]{\begingroup\@sanitize@url \@url }%
\providecommand \@url [1]{\endgroup\@href {#1}{\urlprefix }}%
\providecommand \urlprefix  [0]{URL }%
\providecommand \Eprint [0]{\href }%
\providecommand \doibase [0]{https://doi.org/}%
\providecommand \selectlanguage [0]{\@gobble}%
\providecommand \bibinfo  [0]{\@secondoftwo}%
\providecommand \bibfield  [0]{\@secondoftwo}%
\providecommand \translation [1]{[#1]}%
\providecommand \BibitemOpen [0]{}%
\providecommand \bibitemStop [0]{}%
\providecommand \bibitemNoStop [0]{.\EOS\space}%
\providecommand \EOS [0]{\spacefactor3000\relax}%
\providecommand \BibitemShut  [1]{\csname bibitem#1\endcsname}%
\let\auto@bib@innerbib\@empty
\bibitem [{\citenamefont {Klempt}\ and\ \citenamefont
  {Zaitsev}(2007)}]{Klempt:2007cp}%
  \BibitemOpen
  \bibfield  {author} {\bibinfo {author} {\bibfnamefont {E.}~\bibnamefont
  {Klempt}}\ and\ \bibinfo {author} {\bibfnamefont {A.}~\bibnamefont
  {Zaitsev}},\ }\href {https://doi.org/10.1016/j.physrep.2007.07.006}
  {\bibfield  {journal} {\bibinfo  {journal} {Phys. Rept.}\ }\textbf {\bibinfo
  {volume} {454}},\ \bibinfo {pages} {1} (\bibinfo {year} {2007})},\ \Eprint
  {https://arxiv.org/abs/0708.4016} {arXiv:0708.4016 [hep-ph]} \BibitemShut
  {NoStop}%
\bibitem [{\citenamefont {Crede}\ and\ \citenamefont
  {Meyer}(2009)}]{Crede:2008vw}%
  \BibitemOpen
  \bibfield  {author} {\bibinfo {author} {\bibfnamefont {V.}~\bibnamefont
  {Crede}}\ and\ \bibinfo {author} {\bibfnamefont {C.~A.}\ \bibnamefont
  {Meyer}},\ }\href {https://doi.org/10.1016/j.ppnp.2009.03.001} {\bibfield
  {journal} {\bibinfo  {journal} {Prog. Part. Nucl. Phys.}\ }\textbf {\bibinfo
  {volume} {63}},\ \bibinfo {pages} {74} (\bibinfo {year} {2009})},\ \Eprint
  {https://arxiv.org/abs/0812.0600} {arXiv:0812.0600 [hep-ex]} \BibitemShut
  {NoStop}%
\bibitem [{\citenamefont {Ochs}(2013)}]{Ochs:2013gi}%
  \BibitemOpen
  \bibfield  {author} {\bibinfo {author} {\bibfnamefont {W.}~\bibnamefont
  {Ochs}},\ }\href {https://doi.org/10.1088/0954-3899/40/4/043001} {\bibfield
  {journal} {\bibinfo  {journal} {J. Phys. G}\ }\textbf {\bibinfo {volume}
  {40}},\ \bibinfo {pages} {043001} (\bibinfo {year} {2013})},\ \Eprint
  {https://arxiv.org/abs/1301.5183} {arXiv:1301.5183 [hep-ph]} \BibitemShut
  {NoStop}%
\bibitem [{\citenamefont {Klempt}(2022)}]{Klempt:2022ipu}%
  \BibitemOpen
  \bibfield  {author} {\bibinfo {author} {\bibfnamefont {E.}~\bibnamefont
  {Klempt}},\ }\href@noop {} {\  (\bibinfo {year} {2022})},\ \Eprint
  {https://arxiv.org/abs/2211.12901} {arXiv:2211.12901 [hep-ph]} \BibitemShut
  {NoStop}%
\bibitem [{\citenamefont {Sarantsev}\ \emph {et~al.}(2021)\citenamefont
  {Sarantsev}, \citenamefont {Denisenko}, \citenamefont {Thoma},\ and\
  \citenamefont {Klempt}}]{Sarantsev:2021ein}%
  \BibitemOpen
  \bibfield  {author} {\bibinfo {author} {\bibfnamefont {A.~V.}\ \bibnamefont
  {Sarantsev}}, \bibinfo {author} {\bibfnamefont {I.}~\bibnamefont
  {Denisenko}}, \bibinfo {author} {\bibfnamefont {U.}~\bibnamefont {Thoma}},\
  and\ \bibinfo {author} {\bibfnamefont {E.}~\bibnamefont {Klempt}},\ }\href
  {https://doi.org/10.1016/j.physletb.2021.136227} {\bibfield  {journal}
  {\bibinfo  {journal} {Phys. Lett. B}\ }\textbf {\bibinfo {volume} {816}},\
  \bibinfo {pages} {136227} (\bibinfo {year} {2021})},\ \Eprint
  {https://arxiv.org/abs/2103.09680} {arXiv:2103.09680 [hep-ph]} \BibitemShut
  {NoStop}%
\bibitem [{\citenamefont {Klempt}\ and\ \citenamefont
  {Sarantsev}(2022)}]{Klempt:2021wpg}%
  \BibitemOpen
  \bibfield  {author} {\bibinfo {author} {\bibfnamefont {E.}~\bibnamefont
  {Klempt}}\ and\ \bibinfo {author} {\bibfnamefont {A.~V.}\ \bibnamefont
  {Sarantsev}},\ }\href {https://doi.org/10.1016/j.physletb.2022.136906}
  {\bibfield  {journal} {\bibinfo  {journal} {Phys. Lett. B}\ }\textbf
  {\bibinfo {volume} {826}},\ \bibinfo {pages} {136906} (\bibinfo {year}
  {2022})},\ \Eprint {https://arxiv.org/abs/2112.04348} {arXiv:2112.04348
  [hep-ph]} \BibitemShut {NoStop}%
\bibitem [{\citenamefont {Chen}\ \emph {et~al.}(2023)\citenamefont {Chen},
  \citenamefont {Jiang}, \citenamefont {Chen}, \citenamefont {Liu},
  \citenamefont {Sun},\ and\ \citenamefont {Yang}}]{Chen:2021dvn}%
  \BibitemOpen
  \bibfield  {author} {\bibinfo {author} {\bibfnamefont {F.}~\bibnamefont
  {Chen}}, \bibinfo {author} {\bibfnamefont {X.}~\bibnamefont {Jiang}},
  \bibinfo {author} {\bibfnamefont {Y.}~\bibnamefont {Chen}}, \bibinfo {author}
  {\bibfnamefont {K.-F.}\ \bibnamefont {Liu}}, \bibinfo {author} {\bibfnamefont
  {W.}~\bibnamefont {Sun}},\ and\ \bibinfo {author} {\bibfnamefont {Y.-B.}\
  \bibnamefont {Yang}},\ }\href {https://doi.org/10.1088/1674-1137/accc1c}
  {\bibfield  {journal} {\bibinfo  {journal} {Chin. Phys. C}\ }\textbf
  {\bibinfo {volume} {47}},\ \bibinfo {pages} {063108} (\bibinfo {year}
  {2023})},\ \Eprint {https://arxiv.org/abs/2111.11929} {arXiv:2111.11929
  [hep-lat]} \BibitemShut {NoStop}%
\bibitem [{\citenamefont {Gregory}\ \emph {et~al.}(2012)\citenamefont
  {Gregory}, \citenamefont {Irving}, \citenamefont {Lucini}, \citenamefont
  {McNeile}, \citenamefont {Rago}, \citenamefont {Richards},\ and\
  \citenamefont {Rinaldi}}]{Gregory:2012hu}%
  \BibitemOpen
  \bibfield  {author} {\bibinfo {author} {\bibfnamefont {E.}~\bibnamefont
  {Gregory}}, \bibinfo {author} {\bibfnamefont {A.}~\bibnamefont {Irving}},
  \bibinfo {author} {\bibfnamefont {B.}~\bibnamefont {Lucini}}, \bibinfo
  {author} {\bibfnamefont {C.}~\bibnamefont {McNeile}}, \bibinfo {author}
  {\bibfnamefont {A.}~\bibnamefont {Rago}}, \bibinfo {author} {\bibfnamefont
  {C.}~\bibnamefont {Richards}},\ and\ \bibinfo {author} {\bibfnamefont
  {E.}~\bibnamefont {Rinaldi}},\ }\href
  {https://doi.org/10.1007/JHEP10(2012)170} {\bibfield  {journal} {\bibinfo
  {journal} {JHEP}\ }\textbf {\bibinfo {volume} {10}},\ \bibinfo {pages}
  {170}},\ \Eprint {https://arxiv.org/abs/1208.1858} {arXiv:1208.1858
  [hep-lat]} \BibitemShut {NoStop}%
\bibitem [{\citenamefont {Brett}\ \emph {et~al.}(2020)\citenamefont {Brett},
  \citenamefont {Bulava}, \citenamefont {Darvish}, \citenamefont {Fallica},
  \citenamefont {Hanlon}, \citenamefont {H\"orz},\ and\ \citenamefont
  {Morningstar}}]{Brett:2019tzr}%
  \BibitemOpen
  \bibfield  {author} {\bibinfo {author} {\bibfnamefont {R.}~\bibnamefont
  {Brett}}, \bibinfo {author} {\bibfnamefont {J.}~\bibnamefont {Bulava}},
  \bibinfo {author} {\bibfnamefont {D.}~\bibnamefont {Darvish}}, \bibinfo
  {author} {\bibfnamefont {J.}~\bibnamefont {Fallica}}, \bibinfo {author}
  {\bibfnamefont {A.}~\bibnamefont {Hanlon}}, \bibinfo {author} {\bibfnamefont
  {B.}~\bibnamefont {H\"orz}},\ and\ \bibinfo {author} {\bibfnamefont
  {C.}~\bibnamefont {Morningstar}},\ }\href {https://doi.org/10.1063/5.0008566}
  {\bibfield  {journal} {\bibinfo  {journal} {AIP Conf. Proc.}\ }\textbf
  {\bibinfo {volume} {2249}},\ \bibinfo {pages} {030032} (\bibinfo {year}
  {2020})},\ \Eprint {https://arxiv.org/abs/1909.07306} {arXiv:1909.07306
  [hep-lat]} \BibitemShut {NoStop}%
\bibitem [{\citenamefont {Morningstar}\ and\ \citenamefont
  {Peardon}(1999)}]{Morningstar:1999rf}%
  \BibitemOpen
  \bibfield  {author} {\bibinfo {author} {\bibfnamefont {C.~J.}\ \bibnamefont
  {Morningstar}}\ and\ \bibinfo {author} {\bibfnamefont {M.~J.}\ \bibnamefont
  {Peardon}},\ }\href {https://doi.org/10.1103/PhysRevD.60.034509} {\bibfield
  {journal} {\bibinfo  {journal} {Phys. Rev. D}\ }\textbf {\bibinfo {volume}
  {60}},\ \bibinfo {pages} {034509} (\bibinfo {year} {1999})},\ \Eprint
  {https://arxiv.org/abs/hep-lat/9901004} {arXiv:hep-lat/9901004} \BibitemShut
  {NoStop}%
\bibitem [{\citenamefont {Bali}\ \emph {et~al.}(1993)\citenamefont {Bali},
  \citenamefont {Schilling}, \citenamefont {Hulsebos}, \citenamefont {Irving},
  \citenamefont {Michael},\ and\ \citenamefont {Stephenson}}]{Bali:1993fb}%
  \BibitemOpen
  \bibfield  {author} {\bibinfo {author} {\bibfnamefont {G.~S.}\ \bibnamefont
  {Bali}}, \bibinfo {author} {\bibfnamefont {K.}~\bibnamefont {Schilling}},
  \bibinfo {author} {\bibfnamefont {A.}~\bibnamefont {Hulsebos}}, \bibinfo
  {author} {\bibfnamefont {A.~C.}\ \bibnamefont {Irving}}, \bibinfo {author}
  {\bibfnamefont {C.}~\bibnamefont {Michael}},\ and\ \bibinfo {author}
  {\bibfnamefont {P.~W.}\ \bibnamefont {Stephenson}} (\bibinfo {collaboration}
  {UKQCD}),\ }\href {https://doi.org/10.1016/0370-2693(93)90948-H} {\bibfield
  {journal} {\bibinfo  {journal} {Phys. Lett. B}\ }\textbf {\bibinfo {volume}
  {309}},\ \bibinfo {pages} {378} (\bibinfo {year} {1993})},\ \Eprint
  {https://arxiv.org/abs/hep-lat/9304012} {arXiv:hep-lat/9304012} \BibitemShut
  {NoStop}%
\bibitem [{\citenamefont {Chen}\ \emph {et~al.}(2006)\citenamefont {Chen} \emph
  {et~al.}}]{Chen:2005mg}%
  \BibitemOpen
  \bibfield  {author} {\bibinfo {author} {\bibfnamefont {Y.}~\bibnamefont
  {Chen}} \emph {et~al.},\ }\href {https://doi.org/10.1103/PhysRevD.73.014516}
  {\bibfield  {journal} {\bibinfo  {journal} {Phys. Rev. D}\ }\textbf {\bibinfo
  {volume} {73}},\ \bibinfo {pages} {014516} (\bibinfo {year} {2006})},\
  \Eprint {https://arxiv.org/abs/hep-lat/0510074} {arXiv:hep-lat/0510074}
  \BibitemShut {NoStop}%
\bibitem [{\citenamefont {Athenodorou}\ and\ \citenamefont
  {Teper}(2020)}]{Athenodorou:2020ani}%
  \BibitemOpen
  \bibfield  {author} {\bibinfo {author} {\bibfnamefont {A.}~\bibnamefont
  {Athenodorou}}\ and\ \bibinfo {author} {\bibfnamefont {M.}~\bibnamefont
  {Teper}},\ }\href {https://doi.org/10.1007/JHEP11(2020)172} {\bibfield
  {journal} {\bibinfo  {journal} {JHEP}\ }\textbf {\bibinfo {volume} {11}},\
  \bibinfo {pages} {172}},\ \Eprint {https://arxiv.org/abs/2007.06422}
  {arXiv:2007.06422 [hep-lat]} \BibitemShut {NoStop}%
\bibitem [{\citenamefont {Sakai}\ and\ \citenamefont
  {Sasaki}(2023)}]{Sakai:2022zdc}%
  \BibitemOpen
  \bibfield  {author} {\bibinfo {author} {\bibfnamefont {K.}~\bibnamefont
  {Sakai}}\ and\ \bibinfo {author} {\bibfnamefont {S.}~\bibnamefont {Sasaki}},\
  }\href {https://doi.org/10.1103/PhysRevD.107.034510} {\bibfield  {journal}
  {\bibinfo  {journal} {Phys. Rev. D}\ }\textbf {\bibinfo {volume} {107}},\
  \bibinfo {pages} {034510} (\bibinfo {year} {2023})},\ \Eprint
  {https://arxiv.org/abs/2211.15176} {arXiv:2211.15176 [hep-lat]} \BibitemShut
  {NoStop}%
\bibitem [{\citenamefont {Meyers}\ and\ \citenamefont
  {Swanson}(2013)}]{Meyers:2012ka}%
  \BibitemOpen
  \bibfield  {author} {\bibinfo {author} {\bibfnamefont {J.}~\bibnamefont
  {Meyers}}\ and\ \bibinfo {author} {\bibfnamefont {E.~S.}\ \bibnamefont
  {Swanson}},\ }\href {https://doi.org/10.1103/PhysRevD.87.036009} {\bibfield
  {journal} {\bibinfo  {journal} {Phys. Rev. D}\ }\textbf {\bibinfo {volume}
  {87}},\ \bibinfo {pages} {036009} (\bibinfo {year} {2013})},\ \Eprint
  {https://arxiv.org/abs/1211.4648} {arXiv:1211.4648 [hep-ph]} \BibitemShut
  {NoStop}%
\bibitem [{\citenamefont {Sanchis-Alepuz}\ \emph {et~al.}(2015)\citenamefont
  {Sanchis-Alepuz}, \citenamefont {Fischer}, \citenamefont {Kellermann},\ and\
  \citenamefont {von Smekal}}]{Sanchis-Alepuz:2015hma}%
  \BibitemOpen
  \bibfield  {author} {\bibinfo {author} {\bibfnamefont {H.}~\bibnamefont
  {Sanchis-Alepuz}}, \bibinfo {author} {\bibfnamefont {C.~S.}\ \bibnamefont
  {Fischer}}, \bibinfo {author} {\bibfnamefont {C.}~\bibnamefont
  {Kellermann}},\ and\ \bibinfo {author} {\bibfnamefont {L.}~\bibnamefont {von
  Smekal}},\ }\href {https://doi.org/10.1103/PhysRevD.92.034001} {\bibfield
  {journal} {\bibinfo  {journal} {Phys. Rev. D}\ }\textbf {\bibinfo {volume}
  {92}},\ \bibinfo {pages} {034001} (\bibinfo {year} {2015})},\ \Eprint
  {https://arxiv.org/abs/1503.06051} {arXiv:1503.06051 [hep-ph]} \BibitemShut
  {NoStop}%
\bibitem [{\citenamefont {Souza}\ \emph {et~al.}(2020)\citenamefont {Souza},
  \citenamefont {Narciso~Ferreira}, \citenamefont {Aguilar}, \citenamefont
  {Papavassiliou}, \citenamefont {Roberts},\ and\ \citenamefont
  {Xu}}]{Souza:2019ylx}%
  \BibitemOpen
  \bibfield  {author} {\bibinfo {author} {\bibfnamefont {E.~V.}\ \bibnamefont
  {Souza}}, \bibinfo {author} {\bibfnamefont {M.}~\bibnamefont
  {Narciso~Ferreira}}, \bibinfo {author} {\bibfnamefont {A.~C.}\ \bibnamefont
  {Aguilar}}, \bibinfo {author} {\bibfnamefont {J.}~\bibnamefont
  {Papavassiliou}}, \bibinfo {author} {\bibfnamefont {C.~D.}\ \bibnamefont
  {Roberts}},\ and\ \bibinfo {author} {\bibfnamefont {S.-S.}\ \bibnamefont
  {Xu}},\ }\href {https://doi.org/10.1140/epja/s10050-020-00041-y} {\bibfield
  {journal} {\bibinfo  {journal} {Eur. Phys. J. A}\ }\textbf {\bibinfo {volume}
  {56}},\ \bibinfo {pages} {25} (\bibinfo {year} {2020})},\ \Eprint
  {https://arxiv.org/abs/1909.05875} {arXiv:1909.05875 [nucl-th]} \BibitemShut
  {NoStop}%
\bibitem [{\citenamefont {Kaptari}\ and\ \citenamefont
  {K\"ampfer}(2020)}]{Kaptari:2020qlt}%
  \BibitemOpen
  \bibfield  {author} {\bibinfo {author} {\bibfnamefont {L.~P.}\ \bibnamefont
  {Kaptari}}\ and\ \bibinfo {author} {\bibfnamefont {B.}~\bibnamefont
  {K\"ampfer}},\ }\href {https://doi.org/10.1007/s00601-020-01562-4} {\bibfield
   {journal} {\bibinfo  {journal} {Few Body Syst.}\ }\textbf {\bibinfo {volume}
  {61}},\ \bibinfo {pages} {28} (\bibinfo {year} {2020})},\ \Eprint
  {https://arxiv.org/abs/2004.06523} {arXiv:2004.06523 [hep-ph]} \BibitemShut
  {NoStop}%
\bibitem [{\citenamefont {Huber}\ \emph {et~al.}(2020)\citenamefont {Huber},
  \citenamefont {Fischer},\ and\ \citenamefont
  {Sanchis-Alepuz}}]{Huber:2020ngt}%
  \BibitemOpen
  \bibfield  {author} {\bibinfo {author} {\bibfnamefont {M.~Q.}\ \bibnamefont
  {Huber}}, \bibinfo {author} {\bibfnamefont {C.~S.}\ \bibnamefont {Fischer}},\
  and\ \bibinfo {author} {\bibfnamefont {H.}~\bibnamefont {Sanchis-Alepuz}},\
  }\href {https://doi.org/10.1140/epjc/s10052-020-08649-6} {\bibfield
  {journal} {\bibinfo  {journal} {Eur. Phys. J. C}\ }\textbf {\bibinfo {volume}
  {80}},\ \bibinfo {pages} {1077} (\bibinfo {year} {2020})},\ \Eprint
  {https://arxiv.org/abs/2004.00415} {arXiv:2004.00415 [hep-ph]} \BibitemShut
  {NoStop}%
\bibitem [{\citenamefont {Huber}\ \emph {et~al.}(2021)\citenamefont {Huber},
  \citenamefont {Fischer},\ and\ \citenamefont
  {Sanchis-Alepuz}}]{Huber:2021yfy}%
  \BibitemOpen
  \bibfield  {author} {\bibinfo {author} {\bibfnamefont {M.~Q.}\ \bibnamefont
  {Huber}}, \bibinfo {author} {\bibfnamefont {C.~S.}\ \bibnamefont {Fischer}},\
  and\ \bibinfo {author} {\bibfnamefont {H.}~\bibnamefont {Sanchis-Alepuz}},\
  }\href {https://doi.org/10.1140/epjc/s10052-021-09864-5} {\bibfield
  {journal} {\bibinfo  {journal} {Eur. Phys. J. C}\ }\textbf {\bibinfo {volume}
  {81}},\ \bibinfo {pages} {1083} (\bibinfo {year} {2021})},\ \Eprint
  {https://arxiv.org/abs/2110.09180} {arXiv:2110.09180 [hep-ph]} \BibitemShut
  {NoStop}%
\bibitem [{\citenamefont {Pawlowski}\ \emph {et~al.}(2022)\citenamefont
  {Pawlowski}, \citenamefont {Schneider},\ and\ \citenamefont
  {Wink}}]{Pawlowski:2022oyq}%
  \BibitemOpen
  \bibfield  {author} {\bibinfo {author} {\bibfnamefont {J.~M.}\ \bibnamefont
  {Pawlowski}}, \bibinfo {author} {\bibfnamefont {C.~S.}\ \bibnamefont
  {Schneider}},\ and\ \bibinfo {author} {\bibfnamefont {N.}~\bibnamefont
  {Wink}},\ }\href@noop {} {\  (\bibinfo {year} {2022})},\ \Eprint
  {https://arxiv.org/abs/2202.11123} {arXiv:2202.11123 [hep-th]} \BibitemShut
  {NoStop}%
\bibitem [{\citenamefont {Cuniberti}\ \emph {et~al.}(2001)\citenamefont
  {Cuniberti}, \citenamefont {De~Micheli},\ and\ \citenamefont
  {Viano}}]{Cuniberti:2001hm}%
  \BibitemOpen
  \bibfield  {author} {\bibinfo {author} {\bibfnamefont {G.}~\bibnamefont
  {Cuniberti}}, \bibinfo {author} {\bibfnamefont {E.}~\bibnamefont
  {De~Micheli}},\ and\ \bibinfo {author} {\bibfnamefont {G.~A.}\ \bibnamefont
  {Viano}},\ }\href {https://doi.org/10.1007/s002200000324} {\bibfield
  {journal} {\bibinfo  {journal} {Commun. Math. Phys.}\ }\textbf {\bibinfo
  {volume} {216}},\ \bibinfo {pages} {59} (\bibinfo {year} {2001})},\ \Eprint
  {https://arxiv.org/abs/cond-mat/0109175} {arXiv:cond-mat/0109175}
  \BibitemShut {NoStop}%
\bibitem [{\citenamefont {Burnier}\ \emph {et~al.}(2011)\citenamefont
  {Burnier}, \citenamefont {Laine},\ and\ \citenamefont
  {Mether}}]{Burnier:2011jq}%
  \BibitemOpen
  \bibfield  {author} {\bibinfo {author} {\bibfnamefont {Y.}~\bibnamefont
  {Burnier}}, \bibinfo {author} {\bibfnamefont {M.}~\bibnamefont {Laine}},\
  and\ \bibinfo {author} {\bibfnamefont {L.}~\bibnamefont {Mether}},\ }\href
  {https://doi.org/10.1140/epjc/s10052-011-1619-0} {\bibfield  {journal}
  {\bibinfo  {journal} {Eur. Phys. J. C}\ }\textbf {\bibinfo {volume} {71}},\
  \bibinfo {pages} {1619} (\bibinfo {year} {2011})},\ \Eprint
  {https://arxiv.org/abs/1101.5534} {arXiv:1101.5534 [hep-lat]} \BibitemShut
  {NoStop}%
\bibitem [{\citenamefont {Shi}\ \emph {et~al.}(2023)\citenamefont {Shi},
  \citenamefont {Wang},\ and\ \citenamefont {Zhou}}]{Shi:2022yqw}%
  \BibitemOpen
  \bibfield  {author} {\bibinfo {author} {\bibfnamefont {S.}~\bibnamefont
  {Shi}}, \bibinfo {author} {\bibfnamefont {L.}~\bibnamefont {Wang}},\ and\
  \bibinfo {author} {\bibfnamefont {K.}~\bibnamefont {Zhou}},\ }\href
  {https://doi.org/10.1016/j.cpc.2022.108547} {\bibfield  {journal} {\bibinfo
  {journal} {Comput. Phys. Commun.}\ }\textbf {\bibinfo {volume} {282}},\
  \bibinfo {pages} {108547} (\bibinfo {year} {2023})},\ \Eprint
  {https://arxiv.org/abs/2201.02564} {arXiv:2201.02564 [hep-ph]} \BibitemShut
  {NoStop}%
\bibitem [{\citenamefont {Valentine}\ and\ \citenamefont
  {Sambridge}(2020)}]{valentine2020gaussian}%
  \BibitemOpen
  \bibfield  {author} {\bibinfo {author} {\bibfnamefont {A.~P.}\ \bibnamefont
  {Valentine}}\ and\ \bibinfo {author} {\bibfnamefont {M.}~\bibnamefont
  {Sambridge}},\ }\href {https://doi.org/https://doi.org/10.1093/gji/ggz520}
  {\bibfield  {journal} {\bibinfo  {journal} {Geophysical Journal
  International}\ }\textbf {\bibinfo {volume} {220}},\ \bibinfo {pages} {1632}
  (\bibinfo {year} {2020})}\BibitemShut {NoStop}%
\bibitem [{\citenamefont {Horak}\ \emph {et~al.}(2022)\citenamefont {Horak},
  \citenamefont {Pawlowski}, \citenamefont {Rodr\'\i{}guez-Quintero},
  \citenamefont {Turnwald}, \citenamefont {Urban}, \citenamefont {Wink},\ and\
  \citenamefont {Zafeiropoulos}}]{Horak:2021syv}%
  \BibitemOpen
  \bibfield  {author} {\bibinfo {author} {\bibfnamefont {J.}~\bibnamefont
  {Horak}}, \bibinfo {author} {\bibfnamefont {J.~M.}\ \bibnamefont
  {Pawlowski}}, \bibinfo {author} {\bibfnamefont {J.}~\bibnamefont
  {Rodr\'\i{}guez-Quintero}}, \bibinfo {author} {\bibfnamefont
  {J.}~\bibnamefont {Turnwald}}, \bibinfo {author} {\bibfnamefont {J.~M.}\
  \bibnamefont {Urban}}, \bibinfo {author} {\bibfnamefont {N.}~\bibnamefont
  {Wink}},\ and\ \bibinfo {author} {\bibfnamefont {S.}~\bibnamefont
  {Zafeiropoulos}},\ }\href {https://doi.org/10.1103/PhysRevD.105.036014}
  {\bibfield  {journal} {\bibinfo  {journal} {Phys. Rev. D}\ }\textbf {\bibinfo
  {volume} {105}},\ \bibinfo {pages} {036014} (\bibinfo {year} {2022})},\
  \Eprint {https://arxiv.org/abs/2107.13464} {arXiv:2107.13464 [hep-ph]}
  \BibitemShut {NoStop}%
\bibitem [{\citenamefont {Cyrol}\ \emph {et~al.}(2018)\citenamefont {Cyrol},
  \citenamefont {Pawlowski}, \citenamefont {Rothkopf},\ and\ \citenamefont
  {Wink}}]{Cyrol:2018xeq}%
  \BibitemOpen
  \bibfield  {author} {\bibinfo {author} {\bibfnamefont {A.~K.}\ \bibnamefont
  {Cyrol}}, \bibinfo {author} {\bibfnamefont {J.~M.}\ \bibnamefont
  {Pawlowski}}, \bibinfo {author} {\bibfnamefont {A.}~\bibnamefont
  {Rothkopf}},\ and\ \bibinfo {author} {\bibfnamefont {N.}~\bibnamefont
  {Wink}},\ }\href {https://doi.org/10.21468/SciPostPhys.5.6.065} {\bibfield
  {journal} {\bibinfo  {journal} {SciPost Phys.}\ }\textbf {\bibinfo {volume}
  {5}},\ \bibinfo {pages} {065} (\bibinfo {year} {2018})},\ \Eprint
  {https://arxiv.org/abs/1804.00945} {arXiv:1804.00945 [hep-ph]} \BibitemShut
  {NoStop}%
\bibitem [{\citenamefont {Bonanno}\ \emph {et~al.}(2022)\citenamefont
  {Bonanno}, \citenamefont {Denz}, \citenamefont {Pawlowski},\ and\
  \citenamefont {Reichert}}]{Bonanno:2021squ}%
  \BibitemOpen
  \bibfield  {author} {\bibinfo {author} {\bibfnamefont {A.}~\bibnamefont
  {Bonanno}}, \bibinfo {author} {\bibfnamefont {T.}~\bibnamefont {Denz}},
  \bibinfo {author} {\bibfnamefont {J.~M.}\ \bibnamefont {Pawlowski}},\ and\
  \bibinfo {author} {\bibfnamefont {M.}~\bibnamefont {Reichert}},\ }\href
  {https://doi.org/10.21468/SciPostPhys.12.1.001} {\bibfield  {journal}
  {\bibinfo  {journal} {SciPost Phys.}\ }\textbf {\bibinfo {volume} {12}},\
  \bibinfo {pages} {001} (\bibinfo {year} {2022})},\ \Eprint
  {https://arxiv.org/abs/2102.02217} {arXiv:2102.02217 [hep-th]} \BibitemShut
  {NoStop}%
\bibitem [{\citenamefont {Horak}\ \emph {et~al.}(2021)\citenamefont {Horak},
  \citenamefont {Papavassiliou}, \citenamefont {Pawlowski},\ and\ \citenamefont
  {Wink}}]{Horak:2021pfr}%
  \BibitemOpen
  \bibfield  {author} {\bibinfo {author} {\bibfnamefont {J.}~\bibnamefont
  {Horak}}, \bibinfo {author} {\bibfnamefont {J.}~\bibnamefont
  {Papavassiliou}}, \bibinfo {author} {\bibfnamefont {J.~M.}\ \bibnamefont
  {Pawlowski}},\ and\ \bibinfo {author} {\bibfnamefont {N.}~\bibnamefont
  {Wink}},\ }\bibfield  {journal} {\bibinfo  {journal} {Phys. Rev. D}\ }\textbf
  {\bibinfo {volume} {104}},\ \href
  {https://doi.org/10.1103/PhysRevD.104.074017} {10.1103/PhysRevD.104.074017}
  (\bibinfo {year} {2021}),\ \Eprint {https://arxiv.org/abs/2103.16175}
  {arXiv:2103.16175 [hep-th]} \BibitemShut {NoStop}%
\bibitem [{\citenamefont {Evans}(1992)}]{Evans:1991ky}%
  \BibitemOpen
  \bibfield  {author} {\bibinfo {author} {\bibfnamefont {T.~S.}\ \bibnamefont
  {Evans}},\ }\href {https://doi.org/10.1016/0550-3213(92)90357-H} {\bibfield
  {journal} {\bibinfo  {journal} {Nucl. Phys. B}\ }\textbf {\bibinfo {volume}
  {374}},\ \bibinfo {pages} {340} (\bibinfo {year} {1992})}\BibitemShut
  {NoStop}%
\bibitem [{\citenamefont {Horak}\ \emph {et~al.}(2020)\citenamefont {Horak},
  \citenamefont {Pawlowski},\ and\ \citenamefont {Wink}}]{Horak:2020eng}%
  \BibitemOpen
  \bibfield  {author} {\bibinfo {author} {\bibfnamefont {J.}~\bibnamefont
  {Horak}}, \bibinfo {author} {\bibfnamefont {J.~M.}\ \bibnamefont
  {Pawlowski}},\ and\ \bibinfo {author} {\bibfnamefont {N.}~\bibnamefont
  {Wink}},\ }\href {https://doi.org/10.1103/PhysRevD.102.125016} {\bibfield
  {journal} {\bibinfo  {journal} {Phys. Rev. D}\ }\textbf {\bibinfo {volume}
  {102}},\ \bibinfo {pages} {125016} (\bibinfo {year} {2020})},\ \Eprint
  {https://arxiv.org/abs/2006.09778} {arXiv:2006.09778 [hep-th]} \BibitemShut
  {NoStop}%
\bibitem [{\citenamefont {Dupuis}\ \emph {et~al.}(2021)\citenamefont {Dupuis},
  \citenamefont {Canet}, \citenamefont {Eichhorn}, \citenamefont {Metzner},
  \citenamefont {Pawlowski}, \citenamefont {Tissier},\ and\ \citenamefont
  {Wschebor}}]{Dupuis:2020fhh}%
  \BibitemOpen
  \bibfield  {author} {\bibinfo {author} {\bibfnamefont {N.}~\bibnamefont
  {Dupuis}}, \bibinfo {author} {\bibfnamefont {L.}~\bibnamefont {Canet}},
  \bibinfo {author} {\bibfnamefont {A.}~\bibnamefont {Eichhorn}}, \bibinfo
  {author} {\bibfnamefont {W.}~\bibnamefont {Metzner}}, \bibinfo {author}
  {\bibfnamefont {J.~M.}\ \bibnamefont {Pawlowski}}, \bibinfo {author}
  {\bibfnamefont {M.}~\bibnamefont {Tissier}},\ and\ \bibinfo {author}
  {\bibfnamefont {N.}~\bibnamefont {Wschebor}},\ }\href
  {https://doi.org/10.1016/j.physrep.2021.01.001} {\bibfield  {journal}
  {\bibinfo  {journal} {Phys. Rept.}\ }\textbf {\bibinfo {volume} {910}},\
  \bibinfo {pages} {1} (\bibinfo {year} {2021})},\ \Eprint
  {https://arxiv.org/abs/2006.04853} {arXiv:2006.04853 [cond-mat.stat-mech]}
  \BibitemShut {NoStop}%
\bibitem [{\citenamefont {Cyrol}\ \emph {et~al.}(2016)\citenamefont {Cyrol},
  \citenamefont {Fister}, \citenamefont {Mitter}, \citenamefont {Pawlowski},\
  and\ \citenamefont {Strodthoff}}]{Cyrol:2016tym}%
  \BibitemOpen
  \bibfield  {author} {\bibinfo {author} {\bibfnamefont {A.~K.}\ \bibnamefont
  {Cyrol}}, \bibinfo {author} {\bibfnamefont {L.}~\bibnamefont {Fister}},
  \bibinfo {author} {\bibfnamefont {M.}~\bibnamefont {Mitter}}, \bibinfo
  {author} {\bibfnamefont {J.~M.}\ \bibnamefont {Pawlowski}},\ and\ \bibinfo
  {author} {\bibfnamefont {N.}~\bibnamefont {Strodthoff}},\ }\href
  {https://doi.org/10.1103/PhysRevD.94.054005} {\bibfield  {journal} {\bibinfo
  {journal} {Phys. Rev. D}\ }\textbf {\bibinfo {volume} {94}},\ \bibinfo
  {pages} {054005} (\bibinfo {year} {2016})},\ \Eprint
  {https://arxiv.org/abs/1605.01856} {arXiv:1605.01856 [hep-ph]} \BibitemShut
  {NoStop}%
\bibitem [{\citenamefont {Rasmussen}\ and\ \citenamefont
  {Williams}(2006)}]{Rasmussen:2006gpr}%
  \BibitemOpen
  \bibfield  {author} {\bibinfo {author} {\bibfnamefont {C.}~\bibnamefont
  {Rasmussen}}\ and\ \bibinfo {author} {\bibfnamefont {C.}~\bibnamefont
  {Williams}},\ }\href@noop {} {\emph {\bibinfo {title} {Gaussian Processes for
  Machine Learning}}},\ Adaptive Computation and Machine Learning\ (\bibinfo
  {publisher} {MIT Press},\ \bibinfo {address} {Cambridge, MA, USA},\ \bibinfo
  {year} {2006})\ p.\ \bibinfo {pages} {248}\BibitemShut {NoStop}%
\bibitem [{\citenamefont {Alexandrou}\ \emph {et~al.}(2020)\citenamefont
  {Alexandrou}, \citenamefont {Iannelli}, \citenamefont {Jansen},\ and\
  \citenamefont {Manigrasso}}]{Alexandrou:2020tqq}%
  \BibitemOpen
  \bibfield  {author} {\bibinfo {author} {\bibfnamefont {C.}~\bibnamefont
  {Alexandrou}}, \bibinfo {author} {\bibfnamefont {G.}~\bibnamefont
  {Iannelli}}, \bibinfo {author} {\bibfnamefont {K.}~\bibnamefont {Jansen}},\
  and\ \bibinfo {author} {\bibfnamefont {F.}~\bibnamefont {Manigrasso}}
  (\bibinfo {collaboration} {Extended Twisted Mass}),\ }\href
  {https://doi.org/10.1103/PhysRevD.102.094508} {\bibfield  {journal} {\bibinfo
   {journal} {Phys. Rev. D}\ }\textbf {\bibinfo {volume} {102}},\ \bibinfo
  {pages} {094508} (\bibinfo {year} {2020})},\ \Eprint
  {https://arxiv.org/abs/2007.13800} {arXiv:2007.13800 [hep-lat]} \BibitemShut
  {NoStop}%
\bibitem [{\citenamefont {Del~Debbio}\ \emph {et~al.}(2022)\citenamefont
  {Del~Debbio}, \citenamefont {Giani},\ and\ \citenamefont
  {Wilson}}]{DelDebbio:2021whr}%
  \BibitemOpen
  \bibfield  {author} {\bibinfo {author} {\bibfnamefont {L.}~\bibnamefont
  {Del~Debbio}}, \bibinfo {author} {\bibfnamefont {T.}~\bibnamefont {Giani}},\
  and\ \bibinfo {author} {\bibfnamefont {M.}~\bibnamefont {Wilson}},\ }\href
  {https://doi.org/10.1140/epjc/s10052-022-10297-x} {\bibfield  {journal}
  {\bibinfo  {journal} {Eur. Phys. J. C}\ }\textbf {\bibinfo {volume} {82}},\
  \bibinfo {pages} {330} (\bibinfo {year} {2022})},\ \Eprint
  {https://arxiv.org/abs/2111.05787} {arXiv:2111.05787 [hep-ph]} \BibitemShut
  {NoStop}%
\bibitem [{\citenamefont {Candido}\ \emph {et~al.}(2023)\citenamefont
  {Candido}, \citenamefont {Del~Debbio}, \citenamefont {Giani},\ and\
  \citenamefont {Petrillo}}]{Candido:2023nnb}%
  \BibitemOpen
  \bibfield  {author} {\bibinfo {author} {\bibfnamefont {A.}~\bibnamefont
  {Candido}}, \bibinfo {author} {\bibfnamefont {L.}~\bibnamefont {Del~Debbio}},
  \bibinfo {author} {\bibfnamefont {T.}~\bibnamefont {Giani}},\ and\ \bibinfo
  {author} {\bibfnamefont {G.}~\bibnamefont {Petrillo}},\ }\href
  {https://doi.org/10.22323/1.430.0098} {\bibfield  {journal} {\bibinfo
  {journal} {PoS}\ }\textbf {\bibinfo {volume} {LATTICE2022}},\ \bibinfo
  {pages} {098} (\bibinfo {year} {2023})},\ \Eprint
  {https://arxiv.org/abs/2302.14731} {arXiv:2302.14731 [hep-lat]} \BibitemShut
  {NoStop}%
\bibitem [{\citenamefont {Steinwart}(2002)}]{steinwart:2002}%
  \BibitemOpen
  \bibfield  {author} {\bibinfo {author} {\bibfnamefont {I.}~\bibnamefont
  {Steinwart}},\ }\href {https://dl.acm.org/doi/10.1162/153244302760185252}
  {\bibfield  {journal} {\bibinfo  {journal} {J. Mach. Learn. Res.}\ }\textbf
  {\bibinfo {volume} {2}},\ \bibinfo {pages} {67–93} (\bibinfo {year}
  {2002})}\BibitemShut {NoStop}%
\bibitem [{\citenamefont {Backus}\ and\ \citenamefont
  {Gilbert}(1968)}]{backusgilbert}%
  \BibitemOpen
  \bibfield  {author} {\bibinfo {author} {\bibfnamefont {G.}~\bibnamefont
  {Backus}}\ and\ \bibinfo {author} {\bibfnamefont {F.}~\bibnamefont
  {Gilbert}},\ }\href {https://doi.org/10.1111/j.1365-246X.1968.tb00216.x}
  {\bibfield  {journal} {\bibinfo  {journal} {Geophysical Journal
  International}\ }\textbf {\bibinfo {volume} {16}},\ \bibinfo {pages} {169}
  (\bibinfo {year} {1968})}\BibitemShut {NoStop}%
\bibitem [{\citenamefont {Solak}\ \emph {et~al.}(2003)\citenamefont {Solak},
  \citenamefont {Murray-Smith}, \citenamefont {Leithead}, \citenamefont
  {Leith},\ and\ \citenamefont {Rasmussen}}]{solak:2003}%
  \BibitemOpen
  \bibfield  {author} {\bibinfo {author} {\bibfnamefont {E.}~\bibnamefont
  {Solak}}, \bibinfo {author} {\bibfnamefont {R.}~\bibnamefont {Murray-Smith}},
  \bibinfo {author} {\bibfnamefont {W.}~\bibnamefont {Leithead}}, \bibinfo
  {author} {\bibfnamefont {D.}~\bibnamefont {Leith}},\ and\ \bibinfo {author}
  {\bibfnamefont {C.}~\bibnamefont {Rasmussen}},\ }\href
  {https://proceedings.neurips.cc/paper/2002/file/5b8e4fd39d9786228649a8a8bec4e008-Paper.pdf}
  {\bibfield  {journal} {\bibinfo  {journal} {Advances in Neural Information
  Processing Systems 15}\ ,\ \bibinfo {pages} {1033}} (\bibinfo {year}
  {2003})}\BibitemShut {NoStop}%
\bibitem [{\citenamefont {Agrell}(2019)}]{Agrell:2019}%
  \BibitemOpen
  \bibfield  {author} {\bibinfo {author} {\bibfnamefont {C.}~\bibnamefont
  {Agrell}},\ }\href {http://jmlr.org/papers/v20/19-065.html} {\bibfield
  {journal} {\bibinfo  {journal} {Journal of Machine Learning Research}\
  }\textbf {\bibinfo {volume} {20}},\ \bibinfo {pages} {1} (\bibinfo {year}
  {2019})}\BibitemShut {NoStop}%
\bibitem [{\citenamefont {Ober}\ \emph {et~al.}(2021)\citenamefont {Ober},
  \citenamefont {Rasmussen},\ and\ \citenamefont {van~der
  Wilk}}]{ober2021promises}%
  \BibitemOpen
  \bibfield  {author} {\bibinfo {author} {\bibfnamefont {S.~W.}\ \bibnamefont
  {Ober}}, \bibinfo {author} {\bibfnamefont {C.~E.}\ \bibnamefont
  {Rasmussen}},\ and\ \bibinfo {author} {\bibfnamefont {M.}~\bibnamefont
  {van~der Wilk}},\ }in\ \href
  {https://proceedings.mlr.press/v161/ober21a.html} {\emph {\bibinfo
  {booktitle} {Uncertainty in Artificial Intelligence}}}\ (\bibinfo
  {organization} {PMLR},\ \bibinfo {year} {2021})\ pp.\ \bibinfo {pages}
  {1206--1216}\BibitemShut {NoStop}%
\bibitem [{\citenamefont {Cyrol}\ \emph {et~al.}(2015)\citenamefont {Cyrol},
  \citenamefont {Huber},\ and\ \citenamefont {von Smekal}}]{Cyrol:2014kca}%
  \BibitemOpen
  \bibfield  {author} {\bibinfo {author} {\bibfnamefont {A.~K.}\ \bibnamefont
  {Cyrol}}, \bibinfo {author} {\bibfnamefont {M.~Q.}\ \bibnamefont {Huber}},\
  and\ \bibinfo {author} {\bibfnamefont {L.}~\bibnamefont {von Smekal}},\
  }\href {https://doi.org/10.1140/epjc/s10052-015-3312-1} {\bibfield  {journal}
  {\bibinfo  {journal} {Eur. Phys. J. C}\ }\textbf {\bibinfo {volume} {75}},\
  \bibinfo {pages} {102} (\bibinfo {year} {2015})},\ \Eprint
  {https://arxiv.org/abs/1408.5409} {arXiv:1408.5409 [hep-ph]} \BibitemShut
  {NoStop}%
\bibitem [{\citenamefont {Pawlowski}\ \emph {et~al.}(2023)\citenamefont
  {Pawlowski}, \citenamefont {Schneider},\ and\ \citenamefont
  {Wink}}]{Pawlowski:2021tkk}%
  \BibitemOpen
  \bibfield  {author} {\bibinfo {author} {\bibfnamefont {J.~M.}\ \bibnamefont
  {Pawlowski}}, \bibinfo {author} {\bibfnamefont {C.~S.}\ \bibnamefont
  {Schneider}},\ and\ \bibinfo {author} {\bibfnamefont {N.}~\bibnamefont
  {Wink}},\ }\href {https://doi.org/10.1016/j.cpc.2023.108711} {\bibfield
  {journal} {\bibinfo  {journal} {Comput. Phys. Commun.}\ }\textbf {\bibinfo
  {volume} {287}},\ \bibinfo {pages} {108711} (\bibinfo {year} {2023})},\
  \Eprint {https://arxiv.org/abs/2102.01410} {arXiv:2102.01410 [hep-ph]}
  \BibitemShut {NoStop}%
\bibitem [{\citenamefont {Cyrol}\ \emph {et~al.}(2017)\citenamefont {Cyrol},
  \citenamefont {Mitter},\ and\ \citenamefont {Strodthoff}}]{Cyrol:2016zqb}%
  \BibitemOpen
  \bibfield  {author} {\bibinfo {author} {\bibfnamefont {A.~K.}\ \bibnamefont
  {Cyrol}}, \bibinfo {author} {\bibfnamefont {M.}~\bibnamefont {Mitter}},\ and\
  \bibinfo {author} {\bibfnamefont {N.}~\bibnamefont {Strodthoff}},\ }\href
  {https://doi.org/10.1016/j.cpc.2017.05.024} {\bibfield  {journal} {\bibinfo
  {journal} {Comput. Phys. Commun.}\ }\textbf {\bibinfo {volume} {219}},\
  \bibinfo {pages} {346} (\bibinfo {year} {2017})},\ \Eprint
  {https://arxiv.org/abs/1610.09331} {arXiv:1610.09331 [hep-ph]} \BibitemShut
  {NoStop}%
\bibitem [{\citenamefont {Harris}\ \emph {et~al.}(2020)\citenamefont {Harris},
  \citenamefont {Millman}, \citenamefont {Van Der~Walt}, \citenamefont
  {Gommers}, \citenamefont {Virtanen}, \citenamefont {Cournapeau},
  \citenamefont {Wieser}, \citenamefont {Taylor}, \citenamefont {Berg},
  \citenamefont {Smith} \emph {et~al.}}]{harris2020array}%
  \BibitemOpen
  \bibfield  {author} {\bibinfo {author} {\bibfnamefont {C.~R.}\ \bibnamefont
  {Harris}}, \bibinfo {author} {\bibfnamefont {K.~J.}\ \bibnamefont {Millman}},
  \bibinfo {author} {\bibfnamefont {S.~J.}\ \bibnamefont {Van Der~Walt}},
  \bibinfo {author} {\bibfnamefont {R.}~\bibnamefont {Gommers}}, \bibinfo
  {author} {\bibfnamefont {P.}~\bibnamefont {Virtanen}}, \bibinfo {author}
  {\bibfnamefont {D.}~\bibnamefont {Cournapeau}}, \bibinfo {author}
  {\bibfnamefont {E.}~\bibnamefont {Wieser}}, \bibinfo {author} {\bibfnamefont
  {J.}~\bibnamefont {Taylor}}, \bibinfo {author} {\bibfnamefont
  {S.}~\bibnamefont {Berg}}, \bibinfo {author} {\bibfnamefont {N.~J.}\
  \bibnamefont {Smith}}, \emph {et~al.},\ }\href@noop {} {\bibfield  {journal}
  {\bibinfo  {journal} {Nature}\ }\textbf {\bibinfo {volume} {585}},\ \bibinfo
  {pages} {357} (\bibinfo {year} {2020})}\BibitemShut {NoStop}%
\bibitem [{\citenamefont {Virtanen}\ \emph {et~al.}(2020)\citenamefont
  {Virtanen}, \citenamefont {Gommers}, \citenamefont {Oliphant}, \citenamefont
  {Haberland}, \citenamefont {Reddy}, \citenamefont {Cournapeau}, \citenamefont
  {Burovski}, \citenamefont {Peterson}, \citenamefont {Weckesser},
  \citenamefont {Bright} \emph {et~al.}}]{2020SciPy-NMeth}%
  \BibitemOpen
  \bibfield  {author} {\bibinfo {author} {\bibfnamefont {P.}~\bibnamefont
  {Virtanen}}, \bibinfo {author} {\bibfnamefont {R.}~\bibnamefont {Gommers}},
  \bibinfo {author} {\bibfnamefont {T.~E.}\ \bibnamefont {Oliphant}}, \bibinfo
  {author} {\bibfnamefont {M.}~\bibnamefont {Haberland}}, \bibinfo {author}
  {\bibfnamefont {T.}~\bibnamefont {Reddy}}, \bibinfo {author} {\bibfnamefont
  {D.}~\bibnamefont {Cournapeau}}, \bibinfo {author} {\bibfnamefont
  {E.}~\bibnamefont {Burovski}}, \bibinfo {author} {\bibfnamefont
  {P.}~\bibnamefont {Peterson}}, \bibinfo {author} {\bibfnamefont
  {W.}~\bibnamefont {Weckesser}}, \bibinfo {author} {\bibfnamefont
  {J.}~\bibnamefont {Bright}}, \emph {et~al.},\ }\href@noop {} {\bibfield
  {journal} {\bibinfo  {journal} {Nature methods}\ }\textbf {\bibinfo {volume}
  {17}},\ \bibinfo {pages} {261} (\bibinfo {year} {2020})}\BibitemShut
  {NoStop}%
\end{thebibliography}%

\end{document}